\documentclass[12pt,a4paper]{article}
\pdfoutput=1
\usepackage{amsthm,bm,amsmath,amssymb,epsfig,multicol,multirow}
\usepackage{natbib} 
\usepackage{graphics,rotating,graphicx}
\usepackage{array,multirow,caption}
\usepackage{amssymb}
\usepackage{amsthm}
\usepackage{bm,amsmath,multicol}
\usepackage{booktabs}
\usepackage[colorlinks,bookmarksopen,bookmarksnumbered,citecolor=red,urlcolor=red]{hyperref}
\usepackage{pdflscape}
\usepackage{adjustbox}
\usepackage{tablefootnote}
\usepackage{algorithm}
\usepackage{algpseudocode}

\usepackage{listings}
\usepackage{xcolor}
\usepackage{threeparttable}
\usepackage{booktabs}

\usepackage{float}
\theoremstyle{definition}

\begin{document}

\title{New fast and accurate estimators of the five-parameter lambda distribution (FPLD) with practical applications}

\author{Dilanka S. Dedduwakumara \textsuperscript{a} and Luke A. Prendergast \textsuperscript{b}\\ \\ \\ \textsuperscript{a}School of Public Health, Adelaide University, Australia\\
\textsuperscript{b}Department of Mathematical and Physical Sciences, La Trobe University, Australia}

\maketitle
%\vspace{0.5cm}

\noindent\textbf{ABSTRACT.} When fitting parametric densities, simple distributions are often chosen because they are computationally efficient to estimate and easy to interpret. However, simplicity can result in important features of the data being missed. This paper introduces fast and reliable estimators of the five-parameter lambda distribution (FPLD). The FPLD is highly flexible, can take many different shapes, and provides greater control over skewness and tail behaviour than distributions with fewer parameters. Simulations are provided to demonstrate the excellent performance of the proposed estimators compared with existing methods, at a fraction of the computational cost. Applications in which speed and accuracy are important are also provided, with comparisons to commonly used but potentially overly simplistic approaches. Finally, an R package implementing the proposed estimators and a web application that allows users to determine the FPLD parameters required to approximate well-known distributions are provided.  

\vspace{0.5cm}


\section{Introduction}\label{sec:introduction}

Density estimation is an integral component of statistical analysis, used to describe and investigate distributional properties. Two broad approaches are parametric and non-parametric density estimation. A commonly used non-parametric approach is kernel density estimation, in which the density is estimated by placing a kernel around each observation and combining their contributions across the data space. While this approach offers considerable flexibility, interpretation can be difficult, estimation can be computationally costly, overfitting may identify features that are artifacts of the sample rather than characteristics of the underlying population, and estimates in the tails of the distribution can be unreliable \citep[e.g.][]{silverman1986density}.

In contrast, parametric estimation involves selecting a family of distributions and estimating its parameters from the data. Although this provides a relatively simple and interpretable representation, the selected distribution may be unable to represent important features such as skewness, heavy or light tails, or unusual combinations of these features.  However, flexible distributions that can take many shapes are able to better represent these features at the expense of estimation complexity.  For example, a very common flexible distribution that has four parameters, location, scale, two skewness parameters, is the Generalised Lambda Distribution \citep[GLD][]{freimer1988}.  This distribution can accurately approximate, and for some exactly replicate, many standard distributions (e.g. uniform, exponential, normal, lognormal) as well as many other shapes through different choices of the four parameter settings \citep{dedduwakumara2019simple}.  An even better choice for many applications is the five parameter lambda distribution \citep[FPLD][]{gilchrist2000statistical} where an additional fifth parameter can provide better control of asymmetry and tail behavior for which we provide details of later.  However, introducing the fifth parameter creates an additional computational burden, meaning that existing estimators may take too long to compute when speed is important, or may fail to find an accurate or viable solution.

This paper has several aims.  First, following a brief review of the FPLD in Section \ref{sec:definitions},  we motivate the use of the FPLD by showing that it can provide better density approximations than the GLD, particularly for heavy-tailed and highly skewed distributions. Second, in Section \ref{sec:estimation}, we introduce the new FPLD estimator and review existing estimators. This is followed in Section \ref{sec:simulation} by simulations highlighting the favourable performance and estimation speed of the new estimator. Third, in Section \ref{sec:applications} we provide several practical applications of the FPLD where estimation speed is important. These applications include comparisons with standard approaches that often rely on overly simplistic assumptions. Finally, we provide details of an R package developed to facilitate fitting the FPLD to observed data using the proposed pdQ approach, together with an interactive web application that allows users to obtain the FPLD parameters required to approximate many well-known distributions. We conclude with a discussion of our findings, together with additional supporting details in the Appendix.

\section{Five-parameter lambda distribution}\label{sec:definitions}

Throughout this paper, let $Q(u)=F^{-1}(u)$ denote the quantile function associated with a continuous distribution function $F$ for $u\in(0,1)$, and let $f=F'$ denote the corresponding probability density function, assumed to satisfy $f(x)>0$ throughout its domain. 

\subsection{Definitions}

The five-parameter lambda distribution (FPLD), described by \citet{gilchrist2000statistical}  and subsequently studied by \citet{tarsitano2010comparing} and \cite{mahdi2013estimating}, extends the four-parameter generalized lambda distribution by introducing a fifth parameter that provides additional flexibility in controlling asymmetry and tail behaviour.  \citet{gilchrist2000statistical} expresses the FPLD through the quantile function

\begin{equation}
Q(u)=
\lambda_1+
\frac{\lambda_2}{2}
\left[
(1-\lambda_5)
\left(
\frac{u^{\lambda_3}-1}{\lambda_3}
\right)
-
(1+\lambda_5)
\left(
\frac{(1-u)^{\lambda_4}-1}{\lambda_4}
\right)
\right].
\end{equation}

For consistency with the \texttt{fm5} implementation available in the \texttt{gld} package \citep{Kingt2016gld} for \texttt{R} \citep{R-base}, all estimation procedures developed in this paper use the equivalent inverse-scale representation
\begin{equation}
Q(u)=
\lambda_1+
\frac{1}{\lambda_2}
\left[
(1-\lambda_5)
\left(
\frac{u^{\lambda_3}-1}{\lambda_3}
\right)
-
(1+\lambda_5)
\left(
\frac{(1-u)^{\lambda_4}-1}{\lambda_4}
\right)
\right],
\label{eq:fpld_quantile}
\end{equation}
which differs from the original formulation only in the definition of the scale parameter. When $\lambda_3=0$ or $\lambda_4=0$, the corresponding
terms in \eqref{eq:fpld_quantile} are defined by continuity, with
$(u^{\lambda_3}-1)/\lambda_3 \to \log(u)$ and
$((1-u)^{\lambda_4}-1)/\lambda_4 \to \log(1-u)$, respectively. The four-parameter GLD (FKML parameterisation) can be recovered as a special case by setting $\lambda_5=0$.

The parameters in \eqref{eq:fpld_quantile} have intuitive interpretations. The parameter $\lambda_1$ determines the location of the distribution, $\lambda_2>0$ is an inverse scale parameter, $\lambda_3$ and $\lambda_4$ govern the left and right tail behaviour, respectively, while $\lambda_5\in[-1,1]$ provides additional control over asymmetry by differentially weighting the two sides of the quantile function.

Since the FPLD is specified through its quantile function, its probability density function follows directly from the inverse function theorem. Specifically,
\begin{equation}
f\!\left[Q(u)\right]
=
\frac{1}{q(u)},
\label{eq:fpld_quantile_density}
\end{equation}
where 
$$q(u)=Q'(u)=\frac{1}{\lambda_2}\left[(1-\lambda_5)u^{\lambda_3-1}+
(1+\lambda_5)(1-u)^{\lambda_4-1}
\right]$$ 
denotes the quantile density function \citep{parzen1979,tukey1965} of FPLD.

Substituting this expression into \eqref{eq:fpld_quantile_density} yields the density quantile function of the FPLD,
\begin{equation}
f_Q(u)
=
f\!\left[Q(u)\right]
=
\frac{\lambda_2}
{
(1-\lambda_5)u^{\lambda_3-1}
+
(1+\lambda_5)(1-u)^{\lambda_4-1}
}.
\end{equation}

\subsection{Probability density quantile representation of the FPLD}

The probability density quantile (pdQ), introduced by \citet{staudte2017}, is a location- and scale-invariant representation that characterises the shape of a distribution independently of its location and scale. Since the density quantile function of the FPLD no longer depends on the location parameter $\lambda_1$, but still depends on the inverse scale parameter $\lambda_2$, we obtain the corresponding pdQ by normalising the density quantile function,

\begin{equation}
f_Q^*(u)
=
\frac{f_Q(u)}{\kappa},
\qquad
\kappa
=
\int_0^1
f_Q(u)\,du,
\end{equation}
where $\kappa$ is the normalising constant chosen so that $f_Q^*(u)$ is a probability density on the interval $(0,1)$. Consequently, the pdQ of the FPLD depends only on the three shape parameters $(\lambda_3,\lambda_4,\lambda_5)$ and is invariant to both location and scale.

\subsection{Motivation for the Five-Parameter lambda distribution}\label{sec:motivation}

One of the main motivations for extending the four-parameter GLD to the FPLD is to obtain additional flexibility for modelling asymmetric and heavy–tailed distributions. Although the GLD is already capable of approximating a wide range of distributions \citep{dedduwakumara2019simple}, the additional parameter in the FPLD provides greater flexibility in capturing skewness and tail behaviour. Similar ideas were previously discussed by \cite{tarsitano2010comparing}, where the potential advantages of the FPLD over the GLD were investigated.

\begin{table}[!htbp]
\centering
\caption{Comparison of GLD and FPLD approximations for selected distributions using the pdQ-based estimation approach}
\label{tab:dist_fit}
\hspace*{-0.5cm}
\begin{tabular}{lrrrrrr}
\toprule
Distribution & l1 & l2 & l3 & l4 & l5 & $H^2$\\
\midrule
Uniform(2,5) & 2.0000 & 0.6667 & 1.0000 & 1.0000 & 1.0000 & 0.0000 \\
& 3.5000 & 0.6667 & 1.0000 & 1.0000 &  & 0.0000 \\

Normal (0,1) & 0.0000 & 1.4420 & 0.1469 & 0.1469 & 0.0000 & 0.0001 \\
& 0.0000 & 1.4420 & 0.1469 & 0.1469 & & 0.0001  \\

Lognormal (0,1) & 0.3647 & 2.0114 & 0.2930 & -0.3914 & 0.7789 & 0.0003 \\
& 0.8038 & 1.8141 & 0.7589 & -0.7082 &  & 0.0066 \\

Exponential (1) & 0.0000 & 2.0000 & 1.0000 & 0.0000 & 1.0000 & 0.0000 \\
& 0.0000 & 1.0000 & 30209 & 0.0000 &  & 0.0000\\

$\chi_5^2$ & 3.2113 & 0.5028 & 0.3758 & 0.0246 & 0.3825 & 0.0000 \\
& 4.0559 & 0.4977 & 0.5167 & -0.1470 &  & 0.0016 \\

Pareto (7) & 1.0000 & 14.0000 & 1.0661 & -0.1429 & 1.0000 & 0.0000\\
& 1.0000 & 7.0000 & 19695 & -0.1429 &  & 0.0000\\

Beta (2,3) & 0.3461 & 5.3978 & 0.4453 & 0.3184 & 0.1545 & 0.0001\\
& 0.3770 & 5.3836 & 0.4958 & 0.2637 &  & 0.0004\\

Cauchy (0,1) & 0.0000 & 2.3025 & -0.8418 & -0.8418 & 0.0000 & 0.0003\\
& 0.0000 & 2.3025 & -0.8418 & -0.8418 & & 0.0003\\

Dagum (1,2,3) & 1.3771 & 1.5734 & 0.2450 & -0.4617 & 0.5088 & 0.0000\\
& 1.7568 & 1.5779 & 0.4593 & -0.7716 &  & 0.0028\\

Singh-Maddala (1,2,3) & 0.4457 & 4.5234 & 0.4203 & -0.1022 & 0.1312 & 0.0001\\
& 0.4782 & 4.4903 & 0.4647 & -0.1661 &  & 0.0001\\

\bottomrule
\end{tabular}
\end{table}

To assess the additional flexibility provided by the FPLD, Table ~\ref{tab:dist_fit} compares GLD and FPLD approximations for several common distributions.  The GLD fitting results are taken from \cite{dedduwakumara2019simple}, while the FPLD results are obtained by extending the same pdQ-based estimation approach with quartile matching to include an additional shape parameter, as described in Section \ref{sec:estimation}. For each distribution, the first row corresponds to the FPLD fit and the second row corresponds to the GLD fit. The goodness of fit is evaluated using the squared Hellinger distance ($H^2$) between the target and fitted densities, where smaller values indicate better approximations.  The results show that the FPLD generally provides improved fits compared to the GLD, particularly for skewed and heavy–tailed distributions such as the Lognormal, Dagum, and $\chi^2$ distributions. For symmetric distributions such as the Normal and Uniform distributions, both models perform similarly well. These findings suggest that the additional parameter in the FPLD can provide meaningful improvements in flexibility while still retaining the broad distributional coverage of the GLD family.

\begin{figure}[!htbp]
    \centering
    \includegraphics[scale = 0.20]{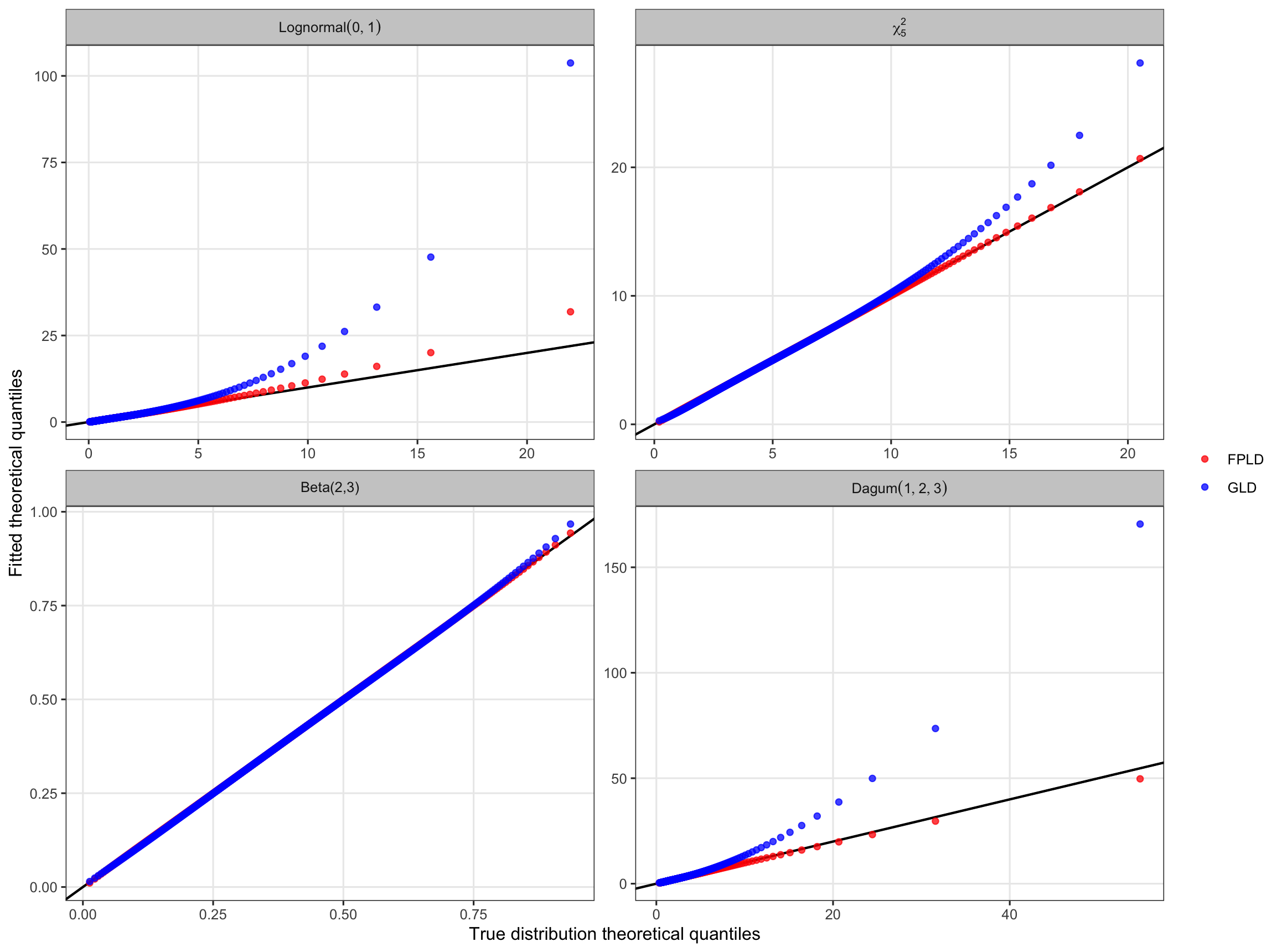}
    \caption{Theoretical QQ comparison of GLD and FPLD fitted distributions against the true theoretical quantiles of the Lognormal$(0,1)$, $\chi^2_5$, Beta$(2,3)$, and Dagum$(1,2,3)$ distributions}
    \label{fig:qqplot}
\end{figure}

To further illustrate the improved flexibility of the FPLD over the GLD, Figure~\ref{fig:qqplot} presents theoretical QQ comparisons between the fitted GLD and FPLD models and the true target distributions.   While both models generally provide satisfactory fits in the central regions, clear differences emerge in the upper tails, particularly for the heavy-tailed Lognormal$(0,1)$ and Dagum$(1,2,3)$ distributions. The GLD shows substantial departures from the reference line in the extreme upper quantiles, whereas the FPLD remains much closer to the target distribution across the tail region. Even for the $\chi^2_5$ and Beta$(2,3)$ distributions, the FPLD exhibits slightly improved tail alignment compared to the GLD.

\begin{table}[!htbp]
\centering
\caption{Upper-tail contributions to the squared Hellinger distance between the true distributions and the fitted GLD and FPLD models, computed over values greater than the true distribution's 75th percentile.}
\label{tab:tail_hellinger}
\begin{tabular}{lcc}
\hline
Distribution & GLD & FPLD \\
\hline
Lognormal$(0,1)$   & 0.0034 & 0.0003 \\
$\chi^2_5$         & 0.0011 & 0.0000 \\
Beta$(2,3)$        & 0.0004 & 0.0000 \\
Dagum$(1,2,3)$     & 0.0015 & 0.0000 \\
\hline
\end{tabular}
\end{table}

To quantify these visual differences, Table~\ref{tab:tail_hellinger} reports the upper-tail contributions to the squared Hellinger distance, calculated over values exceeding the 75th percentile of each true distribution.   Across all considered distributions, the FPLD produces consistently smaller upper-tail Hellinger distances than the GLD, confirming its improved tail representation. The reduction is particularly notable for the Lognormal and Dagum distributions, where tail behaviour is slightly harder to capture using the four-parameter GLD. 

\section{Estimation Methods}\label{sec:estimation}

In this section, we present the proposed pdQ-based estimation methods and briefly outline the existing approaches considered for comparison.  Throughout, let $\mathbf{x}=\{x_i\}_{i=1}^{n}$ denote a random sample of size $n$ from a continuous distribution $F$. The empirical pdQ, denoted by $\hat{f}_Q^*(u)$, is computed as described in \citet{dedduwakumara2021efficient}.

\subsection{pdQ-based estimation method}

The probability density quantile (pdQ) method \citep{dedduwakumara2019simple,dedduwakumara2021efficient} provides a simple two-step procedure for estimating the parameters of the GLD. For this method, the shape parameters are estimated first using the location- and scale-invariant pdQ representation of the distribution. The location and scale parameters are then estimated conditional on the estimated shape parameters using either a quartile matching or a density minimization approach.  In this paper, we extend this method to the FPLD. We also investigate several alternative approaches for estimating the location and scale parameters in addition to those considered by \citet{dedduwakumara2021efficient}. 

\subsubsection{Shape parameter estimation}

Since the pdQ is invariant to location and scale, the initial step of the proposed method is to estimate the shape parameters $\lambda_3$, $\lambda_4$, and $\lambda_5$. These are obtained by minimising the Hellinger distance between the empirical pdQ, $\hat{f}_Q^*(u)$, and the theoretical pdQ, $f_Q^*(u;\lambda_3,\lambda_4,\lambda_5)$.   

Let $\{u_j=(j-1/2)/J\}_{j=1}^{J}$ denote a grid of probability points on $(0,1)$. The shape parameters are estimated by solving

\begin{equation}
(\hat{\lambda}_3, \hat{\lambda}_4, \hat{\lambda}_5)
=
\arg\min_{\lambda_3,\lambda_4,\lambda_5}
\sqrt{
\frac{1}{2J}
\sum_{j=1}^{J}
\left(
\sqrt{\hat{f}_Q^*(u_j)} - \sqrt{f_Q^*(u_j; \lambda_3, \lambda_4, \lambda_5)}
\right)^2
}.
\end{equation}

The above minimisation is performed using \texttt{nlminb} function \citep{gay1990usage} in \texttt{R} \citep{R-base}, which provides a stable and efficient implementation. To obtain suitable starting values for the optimisation, we first evaluate the objective function over a predefined grid of candidate values for the shape parameters. In particular, we consider all possible combinations from  $\lambda_3$, $\lambda_4$ from  -1.5, -1, -0.5, -0.1, 0, 0.1, 0.2, 0.4, 0.8, 1, 1.5 and $\lambda_5$
from -0.5, -0.25, 0, 0.25, 0.5. This grid is adopted from the Starship method implemented for FPLD in the \texttt{gld} package \citep{Kingt2016gld}. In our implementation, we use $J=25$ for smaller samples ($n=100$) and $J=50$ for larger samples ($n\geq 250$), providing a balance between estimation accuracy and computational efficiency.

\subsubsection{Location and scale parameter estimation}

Given the estimated shape parameters $(\hat{\lambda}_3,\hat{\lambda}_4,\hat{\lambda}_5)$, the location and scale parameters $(\hat{\lambda}_1,\hat{\lambda}_2)$ can be  estimated using one of the following approaches.

\subsubsection*{Quartile matching method (pdQ\_qm).}

Let $\hat{x}_p$ denote the empirical $p$th quantile of the observed sample. Given the estimated shape parameters from the first step, the empirical quartiles at probability levels $p\in\{0.25,0.5,0.75\}$ are matched with their corresponding theoretical FPLD quantiles. This gives

\begin{equation}
\hat{x}_p
=
\lambda_1
+
\frac{1}{\lambda_2}
S(p;\hat{\lambda}_3,\hat{\lambda}_4,\hat{\lambda}_5),
\end{equation}
where
\begin{equation}
S(p;\hat{\lambda}_3,\hat{\lambda}_4,\hat{\lambda}_5)
=
(1-\hat{\lambda}_5)
\frac{p^{\hat{\lambda}_3}-1}{\hat{\lambda}_3}
-
(1+\hat{\lambda}_5)
\frac{(1-p)^{\hat{\lambda}_4}-1}{\hat{\lambda}_4}.
\end{equation}

Using the empirical quartiles, a system of linear equations in $\lambda_1$ and $1/\lambda_2$ is obtained. Solving this system yields

\begin{equation}
\hat{\lambda}_2
=
\frac{
S(0.75;\hat{\lambda}_3,\hat{\lambda}_4,\hat{\lambda}_5)
-
S(0.25;\hat{\lambda}_3,\hat{\lambda}_4,\hat{\lambda}_5)
}{
\hat{x}_{0.75}
-
\hat{x}_{0.25}
},
\qquad
\hat{\lambda}_1
=
\hat{x}_{0.5}
-
\frac{
S(0.5;\hat{\lambda}_3,\hat{\lambda}_4,\hat{\lambda}_5)
}{
\hat{\lambda}_2
}.
\end{equation}

\subsubsection*{Density minimisation method (pdQ\_dens)}

In this approach, location and scale parameters $(\lambda_1, \lambda_2)$ are estimated by minimising a Hellinger-type discrepancy between the empirical and fitted FPLD densities evaluated at the observed data points, conditional on the estimated shape parameters $(\hat{\lambda}_3, \hat{\lambda}_4, \hat{\lambda}_5)$.  

Let $\hat f(x)$ denote the empirical density estimated from the observed sample, and let
$f(x;\lambda_1,\lambda_2,\hat\lambda_3,\hat\lambda_4,\hat\lambda_5)$ denote the corresponding FPLD density. The estimators $(\hat{\lambda}_1, \hat{\lambda}_2)$ are obtained as
\begin{equation}
(\hat{\lambda}_1,\hat{\lambda}_2)
=
\arg\min_{\lambda_1,\lambda_2}
\sqrt{
\frac{1}{2n}
\sum_{i=1}^{n}
\left(
\sqrt{\hat f(x_i)}
-
\sqrt{
f(x_i;\lambda_1,\lambda_2,
\hat\lambda_3,\hat\lambda_4,\hat\lambda_5)
}
\right)^2
}.
\end{equation}

The empirical density $\hat{f}(x_i)$ is obtained using the \texttt{demp} function from the R package \texttt{EnvStats} \citep{EnvStats-book}, which employs kernel density estimation. The objective is minimised over $(\lambda_1,\lambda_2)$ using \texttt{nlminb}, with starting values obtained by quartile matching.

\subsubsection*{Method of moments (pdQ\_mom)}

In this approach, location and scale parameters are estimated using a method-of-moments approach based on the mean and variance of the sample. Recall that the FPLD quantile function can be written as
\[
Q(u)=\lambda_1+\frac{1}{\lambda_2}S(u;\lambda_3,\lambda_4,\lambda_5),
\]
where \(S(u;\lambda_3,\lambda_4,\lambda_5)\) denotes the shape-dependent component of the quantile function defined earlier.

The mean of the distribution is therefore
\begin{equation}
\mu
=
\int_0^1 Q(u)\,du
=
\lambda_1+\frac{1}{\lambda_2}S^{(1)},
\end{equation}
where
\[
S^{(1)}
=
\int_0^1
S(u;\hat{\lambda}_3,\hat{\lambda}_4,\hat{\lambda}_5)\,du.
\]

Similarly, the variance is given by
\begin{equation}
\sigma^2
=
\int_0^1
\left\{
Q(u)-\mu
\right\}^2 \,du
=
\frac{1}{\lambda_2^2}
\left[
S^{(2)}
-
\left(S^{(1)}\right)^2
\right],
\end{equation}
where
\[
S^{(2)}
=
\int_0^1
S^2(u;\hat{\lambda}_3,\hat{\lambda}_4,\hat{\lambda}_5)\,du.
\]

Let $\bar{x}$ and $s$ denote the sample mean and standard deviation, respectively. Equating the theoretical and sample moments yields the estimators
\begin{equation}
\hat{\lambda}_2
=
\frac{\sqrt{S^{(2)}-\left(S^{(1)}\right)^2}}{s},
\qquad
\hat{\lambda}_1
=
\bar{x}
-
\frac{S^{(1)}}{\hat{\lambda}_2}.
\end{equation}

Provided that the first two theoretical moments of the FPLD are finite, the integrals $S^{(1)}$ and $S^{(2)}$ are approximated numerically using a fine grid of probability values over $(0,1)$. 

\subsubsection*{Quantile regression methods (pdQ\_qr and pdQ\_wqr)}

Similarly to the quartile matching approach, the following quantile
regression methods exploit the linear relationship between the
empirical and theoretical quantiles implied by the FPLD quantile
function. 

Let
\[
u_i=\frac{i}{n+1}, \qquad i=1,\ldots,n,
\]
denote the plotting positions corresponding to the ordered
observations $x_{(i)}$. Recall that the FPLD quantile function
can be expressed as
\[
Q(u)=\lambda_1+
\frac{1}{\lambda_2}
S(u;\lambda_3,\lambda_4,\lambda_5),
\]
where $S(u;\lambda_3,\lambda_4,\lambda_5)$ is the shape-dependent
component of the quantile function defined earlier.

Conditional on the estimated shape parameters
$(\hat{\lambda}_3,\hat{\lambda}_4,\hat{\lambda}_5)$,
the ordered observations are modelled through the relationship
\begin{equation}
x_{(i)}
=
\lambda_1+
\frac{1}{\lambda_2}
S(u_i;\hat{\lambda}_3,\hat{\lambda}_4,\hat{\lambda}_5)
+\varepsilon_i,
\end{equation}
where $\varepsilon_i$ represents the discrepancy between the
empirical and fitted quantiles.

\paragraph{Unweighted quantile regression (pdQ\_qr).}

The location and scale parameters are estimated by fitting a
linear quantile regression model relating the ordered observations
$x_{(i)}$ to the corresponding theoretical quantiles
$S(u_i;\hat{\lambda}_3,\hat{\lambda}_4,\hat{\lambda}_5)$.
Specifically, the regression coefficients are obtained by solving
\begin{equation}
(\hat{\beta}_0,\hat{\beta}_1)
=
\arg\min_{\beta_0,\beta_1}
\sum_{i=1}^{n}
\rho_{\tau}
\left[
x_{(i)}
-
\beta_0
-
\beta_1
S(u_i;\hat{\lambda}_3,\hat{\lambda}_4,\hat{\lambda}_5)
\right],
\end{equation}
where
\[
\rho_{\tau}(r)
=
r\left\{\tau-\mathbf{1}(r<0)\right\}
\]
is the quantile regression loss function, $r$ denotes a residual,
and $\tau\in(0,1)$ is the quantile level.

The estimators of the location and scale parameters are then
given by
\begin{equation}
\hat{\lambda}_1=\hat{\beta}_0,
\qquad
\hat{\lambda}_2=\frac{1}{\hat{\beta}_1}.
\end{equation}

\paragraph{Weighted quantile regression (pdQ\_wqr).}

Although the unweighted quantile regression method assigns equal
importance to all ordered observations, the sampling variability
of empirical quantiles generally differs across probability levels.
In particular, under standard regularity conditions, the
asymptotic variance of the empirical quantile at probability
level $u$ is given by \citep{ferguson1996course,dasgupta2008asymptotic,serfling2009approximation}
\begin{equation}
\operatorname{Var}\left[\hat{Q}(u)\right]
\approx
\frac{u(1-u)}
{n f^2\left[Q(u)\right]},
\end{equation}
where $f\left[Q(u)\right]$ denotes the probability density function
evaluated at the population quantile $Q(u)$.

Consequently, the corresponding asymptotic standard error is
\begin{equation}
\operatorname{SE}\left[\hat{Q}(u)\right]
\approx
\frac{\sqrt{u(1-u)}}
{\sqrt{n}\,f\left[Q(u)\right]}.
\end{equation}

Motivated by this expression, we introduce a weighted quantile
regression estimator in which the weights are inversely
proportional to the asymptotic standard errors of the empirical
quantiles. Since the common factor $\sqrt{n}$ does not affect
the minimisation, the weights are defined as
\begin{equation}
w_i
=
\frac{
f_0\left[
S(u_i;\hat{\lambda}_3,\hat{\lambda}_4,\hat{\lambda}_5)
\right]
}{
\sqrt{u_i(1-u_i)}
},
\qquad i=1,\ldots,n,
\end{equation}
where $f_0$ denotes the density of the standardised FPLD
with location parameter zero, scale parameter one, and
shape parameters
$(\hat{\lambda}_3,\hat{\lambda}_4,\hat{\lambda}_5)$.

The weighted quantile regression estimates are then obtained
by minimising
\begin{equation}
(\hat{\beta}_0^{\,w},\hat{\beta}_1^{\,w})
=
\arg\min_{\beta_0,\beta_1}
\sum_{i=1}^{n}
w_i
\rho_{\tau}
\left[
x_{(i)}
-
\beta_0
-
\beta_1
S(u_i;\hat{\lambda}_3,\hat{\lambda}_4,\hat{\lambda}_5)
\right].
\end{equation}

The corresponding location and scale estimators are
\begin{equation}
\hat{\lambda}_1^{\,w}
=
\hat{\beta}_0^{\,w},
\qquad
\hat{\lambda}_2^{\,w}
=
\frac{1}{\hat{\beta}_1^{\,w}}.
\end{equation}

The weighting scheme reduces the relative contribution of
quantiles with larger estimated sampling variability, while
retaining the computational simplicity of linear quantile
regression. The unweighted and weighted formulations are
evaluated using $\tau=0.5$, corresponding to median regression.

\subsection{Existing estimation methods for FPLD}

In addition to the proposed pdQ-based methods, several existing approaches can be used for estimating the parameters of the FPLD. In this section, we briefly describe the Starship, Maximum Likelihood Estimation (MLE), Quantile Matching Estimation (QME), and Maximum Goodness-of-fit Estimation (MGE) methods.

For consistency, all estimation procedures, except Starship, were implemented using custom \textsf{R} functions developed specifically for this study. The same grid over the shape parameters $(\lambda_3,\lambda_4,\lambda_5)$ used in the
pdQ-based methods is employed to obtain suitable starting values. For each candidate combination, the corresponding location and scale parameters $(\lambda_1,\lambda_2)$ are obtained by matching the median and interquartile range, respectively. The full parameter vector is then jointly optimised using the corresponding method-specific
objective function and the \texttt{nlminb} routine in
\textsf{R}. In contrast, the Starship estimator was obtained from the \texttt{gld} package \citep{Kingt2016gld} in \textsf{R}.

\subsubsection*{Starship method}

The Starship method, originally introduced by \cite{owen1988starship}, is a computationally intensive approach that focuses on assessing the fit to a base distribution through an inverse transformation. This idea was later adapted by \cite{king1999starship} for the FPLD, where the distribution function is obtained numerically and the parameter values are selected to minimise the deviation from a uniform distribution.

In the context of the FPLD model, this is achieved by transforming the observed data through the fitted cumulative distribution function and evaluating the goodness-of-fit to the uniform distribution using an Anderson-Darling type criterion. A grid of candidate values for the shape parameters is considered, and for each combination, the corresponding location and scale parameters $(\lambda_1, \lambda_2)$ are obtained via quartile matching. The parameter set that minimises the goodness-of-fit criterion is selected as the final estimate.

While the Starship method provides excellent estimates across a wide range of distributions, it is computationally demanding, particularly for large sample sizes \citep{dedduwakumara2021efficient,vandeskog2022quantile}.

\subsubsection*{Maximum likelihood estimation (MLE)}

The MLE approach estimates the parameters by maximising the likelihood of the observed data under the FPLD model. Given observations $x_1,\ldots,x_n$, the log-likelihood function is defined as

\begin{equation}
\ell(\boldsymbol{\lambda})
=
\sum_{i=1}^{n}
\log f(x_i;\boldsymbol{\lambda}),
\end{equation}
where $\boldsymbol{\lambda}=(\lambda_1,\lambda_2,\lambda_3,\lambda_4,\lambda_5)$ denotes the vector of FPLD parameters and $f(x;\boldsymbol{\lambda})$ is the FPLD density function defined in Section ~\ref{sec:definitions}. As noted by \citet{tarsitano2010comparing,tarsitano2005estimation}, maximum likelihood estimation is also computationally demanding due to the complexity of the likelihood surface in practice.

\subsubsection*{Quantile matching estimation (QME)}

The QME method estimates the parameters by matching empirical quantiles with their corresponding theoretical quantiles. Motivated by the percentile matching approach of \citet{karian1999fitting,tarsitano2005estimation}, the parameter estimates are obtained by minimising
\begin{equation}
\sum_{j=1}^{5}
\left(
\hat{x}_{p_j}
-
Q(p_j;\boldsymbol{\lambda})
\right)^2,
\qquad
p_j\in\{0.10,\,0.25,\,0.50,\,0.75,\,0.90\},
\end{equation}
where $\hat{x}_{p_j}$ and $Q(p_j;\boldsymbol{\lambda})$ denote the empirical and theoretical quantiles corresponding to probability level $p_j$, respectively. These probability levels provide information from both the central region and the tails of the distribution while avoiding the instability associated with more extreme sample quantiles. 

\subsubsection*{Maximum goodness-of-fit estimation (MGE)}

The MGE method estimates the FPLD parameters by minimising the distance between the empirical distribution function and the fitted theoretical distribution function. Several goodness-of-fit criteria may be used for this purpose, including the Kolmogorov--Smirnov, Anderson--Darling, and Cramér--von Mises statistics \citep{Stephens1986EDF}. In this study, we use the Cramér--von Mises criterion, which is also the default goodness-of-fit measure in the \texttt{fitdistrplus} package \citep{delignette2015fitdistrplus} .

Specifically for ordered observations $x_{(1)},\ldots,x_{(n)}$ the FPLD parameter estimates are obtained by minimising
\begin{equation}
\frac{1}{12n^2}
+
\frac{1}{n}
\sum_{i=1}^{n}
\left[
F\!\left(x_{(i)};\boldsymbol{\lambda}\right)
-
\frac{(2i-1)}{2n}
\right]^2,
\end{equation}
where 
$F(x;\boldsymbol{\lambda})$ denotes the FPLD cumulative distribution function. 

\section{Simulation Results}\label{sec:simulation}

In this section, we present simulation studies to evaluate the performance of the proposed pdQ-based estimation methods in comparison with existing estimation approaches for FPLD presented in Section ~\ref{sec:estimation}.

\subsection{Comparison of Estimation Methods}

We first compare the average Jensen--Shannon divergence (JSD) between the
empirical density and the fitted FPLD density over 500 simulated data sets.
The JSD provides a symmetric and bounded measure of the discrepancy between
two probability distributions, making it a suitable criterion for comparing
the fitted densities obtained by the competing estimators
\citep{lin2002divergence,endres2003new}.

For each simulated data set, a sample-point-based JSD was computed by evaluating the empirical density $\hat f$ and the fitted FPLD density $f_{\hat{\boldsymbol{\lambda}}}$ at the observed data points $x_1,\ldots,x_n$. The resulting density values were normalised to obtain probability vectors
\[
p_i=\frac{\hat f(x_i)}{\sum_{j=1}^{n}\hat f(x_j)},
\qquad
q_i=\frac{f_{\hat{\boldsymbol{\lambda}}}(x_i)}
{\sum_{j=1}^{n}f_{\hat{\boldsymbol{\lambda}}}(x_j)}.
\]
The discrete Jensen--Shannon divergence was then calculated as
\[
\operatorname{JSD}(P,Q)
=\frac12\sum_{i=1}^{n}p_i\log_2\left(\frac{2p_i}{p_i+q_i}\right)
+\frac12\sum_{i=1}^{n}q_i\log_2\left(\frac{2q_i}{p_i+q_i}\right).
\]
The empirical density was estimated using the \texttt{demp} function from the R package \texttt{EnvStats} \citep{EnvStats-book}, and the JSD was computed using the \texttt{philentropy} \citep{drost2018philentropy} package with base-2 logarithms.

The simulation study considers a diverse range of distributions, including
the Normal and Lognormal distributions with $\mu=0$ and $\sigma=1$, the
Singh--Maddala distribution with parameters $a=1.6971$, $b=87.6981$, and
$q=8.3679$ \citep{mcdonald1984some}, and the Dagum distribution with
parameters $a=4.273$, $b=14.28$, and $p=0.36$
\citep{kleiber2008guide}. In addition, the Exponential, $\chi^2_{5}$, and
Beta$(2,3)$ distributions are included to represent a broad range of
distributional shapes. Results are reported for sample sizes $n=100$, $n=250$, $n=500$,
and $n=1000$. The main text focuses on the two best-performing pdQ estimators,
pdQ\_dens and pdQ\_wqr ($\tau = 0.50$). The results for the remaining
estimators are provided in Appendix~\ref{app:simulation_results}.

\begin{table}[H]
\centering
\caption{This table shows a comparison of the Average Jensen-Shannon divergence (JSD) between the empirical density and the fitted FPLD density over 500 simulation replications. The considered methods are Methods of Probability Density Quantiles (pdQ's), Maximum Likelihood Estimation (MLE), Quantile Matching Estimation (QME), Maximum Goodness-of-Fit Estimation (MGE; minimizing the Cram\'er--von Mises distance) and Starship estimation. Smaller JSD values indicate better performance.}
\label{tab:JSD}
\begin{threeparttable}
\begin{adjustbox}{max width=\textwidth}
\footnotesize
\begin{tabular}{cccccccc}
\toprule
Distribution & $n$ & pdQ\_dens & pdQ\_wqr & MLE & QME & MGE & Starship\\
\midrule

Normal(0,1)
&100  & \textbf{0.0008} & 0.0010 & 0.0010 & 0.0067 & 0.0019 & \textbf{0.0008}\\
&250  & \textbf{0.0003} & 0.0005 & 0.0006 & 0.0038 & 0.0007 & 0.0005\\
&500  & \textbf{0.0002} & 0.0003 & 0.0004 & 0.0023 & 0.0004 & 0.0003\\
&1000 & \textbf{0.0002} & \textbf{0.0002} & 0.0003 & 0.0013 & \textbf{0.0002} & \textbf{0.0002}\\

\midrule

Lognormal(0,1)
&100  & \textbf{0.0036} & 0.0123 & 0.0098 & 0.0242 & 0.0143 & 0.0074\\
&250  & \textbf{0.0023} & 0.0069 & 0.0055 & 0.0214 & 0.0094 & 0.0045\\
&500  & \textbf{0.0018} & 0.0049 & 0.0039 & 0.0170 & 0.0060 & 0.0033\\
&1000 & \textbf{0.0013} & 0.0034 & 0.0028 & 0.0140 & 0.0040 & 0.0024\\

\midrule

Exponential(1)
&100  & \textbf{0.0065} & 0.0219 & 0.0143 & 0.0257 & 0.0175 & 0.0119\\
&250  & \textbf{0.0056} & 0.0154 & 0.0103 & 0.0177 & 0.0126 & 0.0097\\
&500  & \textbf{0.0051} & 0.0119 & 0.0086 & 0.0154 & 0.0101 & 0.0082\\
&1000 & \textbf{0.0045} & 0.0088 & 0.0076 & 0.0136 & 0.0084 & 0.0071\\

\midrule

Dagum
&100  & 0.0022 & 0.0025 & 0.0023 & 0.0078 & 0.0025 & \textbf{0.0011}\\
&250  & 0.0007 & 0.0010 & 0.0008 & 0.0050 & 0.0010 & \textbf{0.0005}\\
&500  & 0.0005 & 0.0006 & 0.0005 & 0.0030 & 0.0006 & \textbf{0.0003}\\
&1000 & 0.0003 & 0.0004 & 0.0003 & 0.0018 & 0.0003 & \textbf{0.0002}\\

\midrule

Sing-Maddala
&100  & 0.0021 & 0.0035 & 0.0033 & 0.0111 & 0.0041 & \textbf{0.0016}\\
&250  & 0.0009 & 0.0015 & 0.0011 & 0.0080 & 0.0018 & \textbf{0.0008}\\
&500  & 0.0006 & 0.0010 & 0.0007 & 0.0054 & 0.0010 & \textbf{0.0005}\\
&1000 & 0.0004 & 0.0005 & 0.0004 & 0.0036 & 0.0006 & \textbf{0.0003}\\

\midrule

$\chi^2_{5}$
&100  & \textbf{0.0016} & 0.0028 & 0.0028 & 0.0112 & 0.0040 & \textbf{0.0016}\\
&250  & \textbf{0.0008} & 0.0014 & 0.0010 & 0.0078 & 0.0019 & 0.0009\\
&500  & \textbf{0.0005} & 0.0009 & 0.0007 & 0.0057 & 0.0011 & 0.0006\\
&1000 & \textbf{0.0003} & 0.0005 & 0.0004 & 0.0039 & 0.0006 & 0.0004\\

\midrule

Beta(2,3)
&100  & 0.0019 & 0.0018 & 0.0021 & 0.0092 & 0.0032 & \textbf{0.0009}\\
&250  & 0.0008 & 0.0008 & 0.0008 & 0.0061 & 0.0013 & \textbf{0.0005}\\
&500  & 0.0005 & 0.0005 & 0.0005 & 0.0039 & 0.0007 & \textbf{0.0003}\\
&1000 & \textbf{0.0003} & \textbf{0.0003} & \textbf{0.0003} & 0.0026 & 0.0004 & \textbf{0.0003}\\

\bottomrule
\end{tabular}
\end{adjustbox}

\begin{tablenotes}
\scriptsize
\item \textit{Notes:} Bold values indicate the lowest JSD value.
\end{tablenotes}

\end{threeparttable}
\end{table}

Table~\ref{tab:JSD} shows that pdQ\_dens performs particularly well, achieving the smallest or jointly smallest JSD values for the Normal, Lognormal, Exponential, and $\chi^2_5$ distributions across all four sample sizes. For the Dagum, Singh--Maddala, and Beta$(2,3)$ distributions, Starship performs slightly better, although pdQ\_dens remains competitive. Overall, pdQ\_dens demonstrates strong density-fitting performance across the distributions considered. While pdQ\_wqr generally produces larger JSD values than pdQ\_dens, its performance remains competitive, with the additional advantage of lower computational cost, as demonstrated later.

For most methods and distributions, JSD decreases with increasing sample size, while the relative performance of the estimators remains broadly consistent.

\begin{table}[H]
\centering
\caption{Comparison of upper-tail Jensen--Shannon divergence (JSD) between the empirical upper-tail density and the fitted FPLD upper-tail density for $n=500$ over 500 simulation replications. The upper tail is defined as observations exceeding the empirical 75th percentile. Smaller JSD values indicate better upper-tail fit.}
\label{tab:tail_jsd}
\begin{threeparttable}
\begin{adjustbox}{max width=\textwidth}
\footnotesize
\begin{tabular}{lcccccc}
\toprule
Distribution & pdQ\_dens & pdQ\_wqr & MLE & QME & MGE & Starship \\
\midrule
Normal(0,1)      & \textbf{0.0079} & 0.0090 & 0.0087 & 0.0208 & 0.0108 & 0.0093\\
Lognormal(0,1)   & \textbf{0.0151} & 0.0209 & 0.0230 & 0.0271 & 0.0229 & 0.0224 \\
Exponential(1)   & \textbf{0.0113} & 0.0160 & 0.0160 & 0.0222 & 0.0160 & 0.0152 \\
Dagum            & \textbf{0.0115} & 0.0135 & 0.0131 & 0.0217 & 0.0149 & 0.0135 \\
Sing-Maddala     & \textbf{0.0094} & 0.0117 & 0.0116 & 0.0211 & 0.0133 & 0.0119 \\
$\chi^2_{5}$     & \textbf{0.0092} & 0.0116 & 0.0118 & 0.0194 & 0.0131 & 0.0119\\
Beta(2,3)    & \textbf{0.0064} & 0.0073 & 0.0067 & 0.0205 & 0.0097 & 0.0077\\
\bottomrule
\end{tabular}
\end{adjustbox}
\begin{tablenotes}
\scriptsize
\item \textit{Notes:} Bold values indicate the lowest upper-tail JSD for each distribution.
\end{tablenotes}
\end{threeparttable}
\end{table}

While Table~\ref{tab:JSD} assesses the overall density approximation,
Table~\ref{tab:tail_jsd} focuses on the upper tail, defined by observations
exceeding the empirical 75th percentile. Since the overall patterns were similar across sample sizes, results are presented for $n=500$ only.  The empirical density was estimated using only the upper-tail observations, while the fitted FPLD density was evaluated at the same points using parameter estimates obtained from the full sample. Both sets of density evaluations were then separately normalised to sum to one over the upper-tail observations, and the JSD was computed as described above. This provides a sample-point-based measure of the agreement between the empirical and fitted densities within the upper tail.

The upper-tail results are consistent with the full-density comparison, with pdQ\_dens achieving the smallest JSD for all the distributions. The competing methods also perform well in several cases, although their JSD values are generally larger than those of pdQ\_dens. Overall, the results show that the strong performance of pdQ\_dens is maintained when the comparison is restricted to the upper tail.

To further investigate the performance of the estimators under increasing skewness, an additional simulation study was conducted using Lognormal distributions with $\mu=0$ and
$\sigma \in \{0.5,1,1.5,2\}$. For each value of $\sigma$, 500 samples of size $n=250$ were generated, and the average JSD was calculated for each estimation method.

\begin{figure}[htbp]
    \centering
    \includegraphics[scale=0.2]{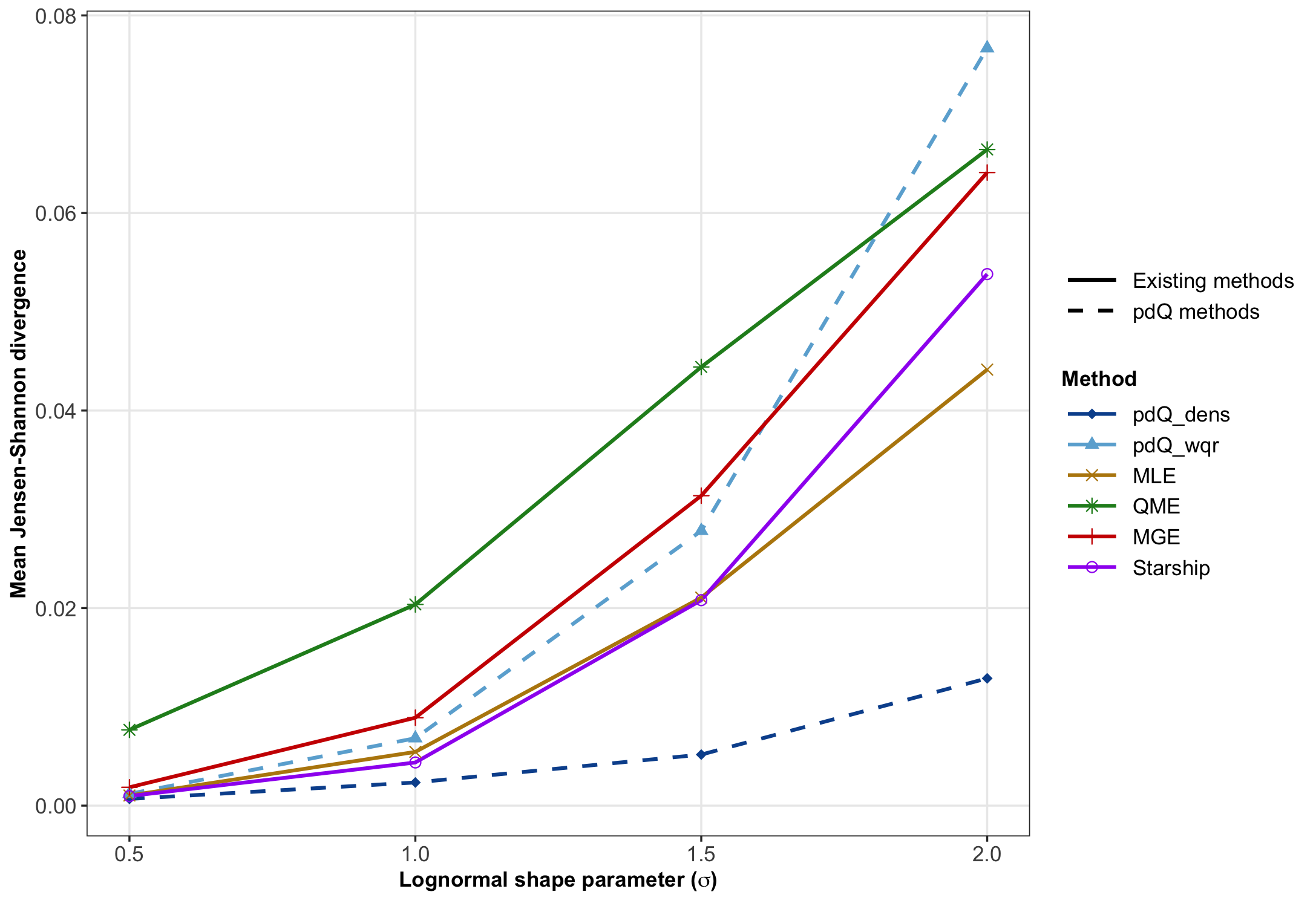}
    \caption{Average Jensen--Shannon divergence of the competing
    FPLD estimation methods for Lognormal distributions with
    increasing shape parameter $\sigma$, based on 500 simulated
    samples of size $n=250$.}
    \label{fig:skew_jsd}
\end{figure}

Figure~\ref{fig:skew_jsd} shows that the average JSD increases with $\sigma$ for all methods, although the increase varies considerably between estimators. The pdQ\_dens method performs particularly well and gives the smallest JSD across all values of $\sigma$. Its advantage over the other methods becomes more noticeable as skewness increases. While pdQ\_wqr also performs well at lower levels of skewness, its performance becomes poorer as $\sigma$ increases. At $\sigma=2$, it gives the largest JSD among the methods considered. Overall, pdQ\_dens continues to perform well even for highly skewed Lognormal distributions.

In addition to estimation accuracy, computational efficiency is an important consideration, particularly when repeated model fitting is required.

\begin{figure}[h]
    \centering
    \includegraphics[scale=0.17]{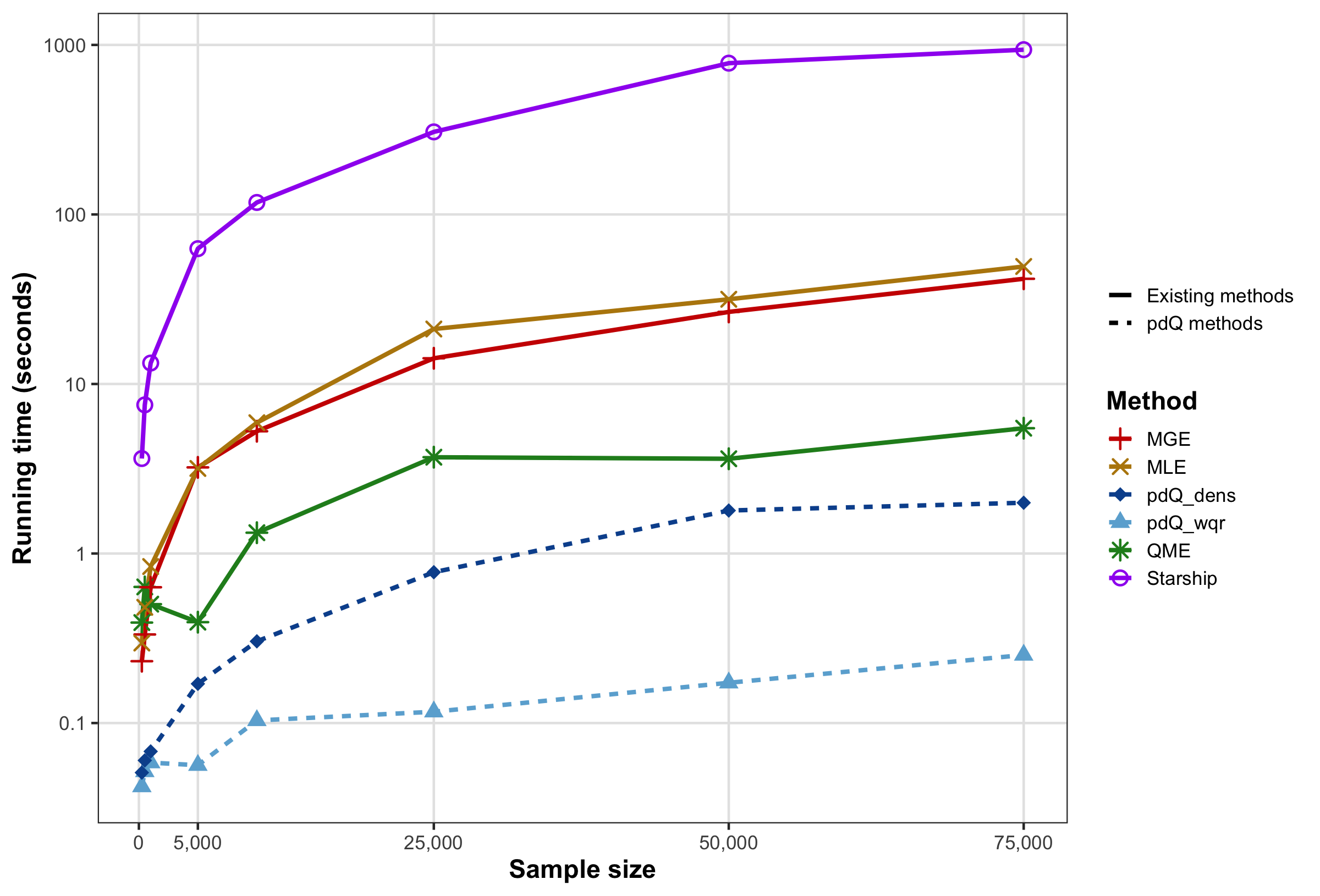}
    \caption{Average running times of the competing estimation methods across different sample sizes, shown on a logarithmic scale. Data were generated from a
    five-parameter GLD with $\lambda_1=0$, $\lambda_2=1.4420$,
    $\lambda_3=0.1469$, $\lambda_4=0.1469$, and $\lambda_5=0$.
    Each running time is the average over five simulation replications.}
    \label{fig:run_times}
\end{figure}

Figure~\ref{fig:run_times} compares the average running times of the competing estimators across different sample sizes. The running times were measured on a 2023 Apple iMac with an Apple M3 chip (8-core CPU) and 16 GB of memory. Both pdQ estimators considered here are considerably faster than the existing methods. Among the two, pdQ\_wqr is the fastest, while pdQ\_dens requires more computation due to the density-based optimisation of the location and scale parameters. Nevertheless, pdQ\_dens remains substantially faster than the existing methods.

Among the existing methods, QME is generally the fastest, whereas Starship requires the most computation. The difference becomes more noticeable as the sample size increases. For example, at $n=75{,}000$, the average running times of pdQ\_wqr and Starship are 0.252 and 937.373 seconds, respectively, making Starship approximately 3,720 times slower. Overall, pdQ\_dens provides the best density-fitting performance, while pdQ\_wqr offers the greatest computational efficiency. This makes the proposed methods useful for applications involving repeated fitting, such as bootstrap procedures and rolling-window analyses, which are considered in Section~\ref{sec:applications}.

\section{Applications}\label{sec:applications}

We illustrate the practical performance of the proposed estimators through three real-data applications. These include a comparison with existing FPLD estimation methods, rolling distributional modelling, and bootstrap inference.

\subsection{Comparison of estimation methods for diurnal temperature range data}
\label{sec:dtr_application}

\begin{figure}[htbp]
    \centering
    \includegraphics[width=0.9\textwidth]{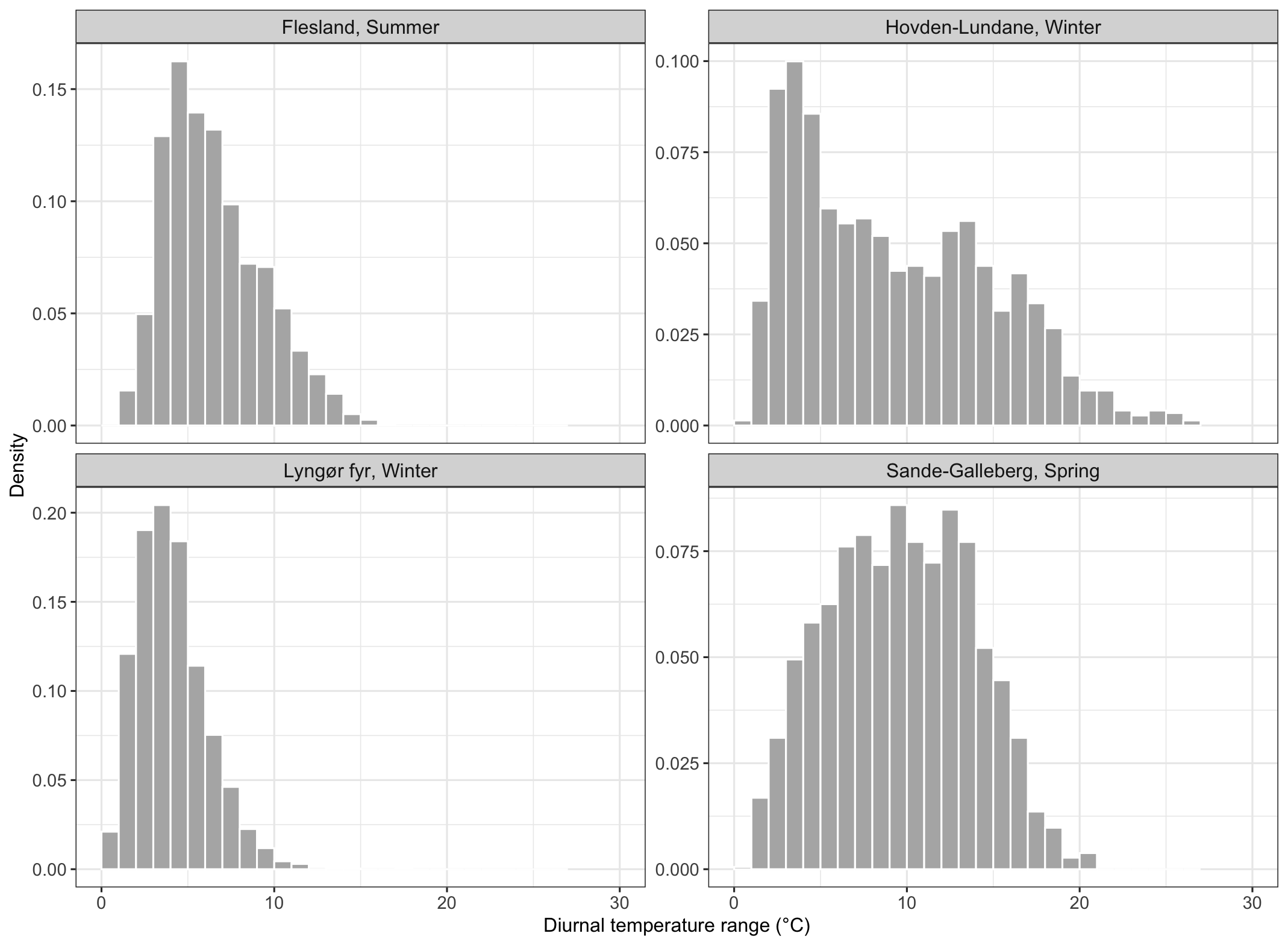}
    \caption{Histograms of diurnal temperature range for the four selected weather
    stations and seasons.}
    \label{fig:dtr_hist}
\end{figure}

We first compare the proposed estimators with existing methods using diurnal temperature range (DTR) data. DTR is defined as the difference between the daily maximum and minimum temperatures and has previously been modelled using the FPLD by \cite{vandeskog2022quantile}. The data were obtained from the publicly available GitHub repository associated with that study (\url{https://github.com/siliusmv/FPLD}).
We consider four station--season combinations examined in the study: Flesland in
summer, Hovden-Lundane in winter, Lyng{\o}r fyr in winter, and Sande-Galleberg in
spring, with sample sizes 2759, 1461, 2707, and 1840, respectively. Figure~\ref{fig:dtr_hist} shows that these datasets have different distributional shapes, making them useful for comparing the estimation methods.

For each dataset, we fitted the FPLD using the proposed and existing estimation methods. As in the simulation study, density-fitting performance was assessed using the JSD. We also recorded the computation times, with each value averaged over five runs.

\begin{table}[H]
\centering
\caption{Jensen-Shannon divergence (JSD) and mean computation time (seconds) for FPLD estimates fitted to the diurnal temperature range datasets. Smaller JSD values indicate better fit.}
\label{tab:dtr_comparison}

\begin{threeparttable}
\begin{adjustbox}{max width=\textwidth}
\footnotesize

\begin{tabular}{llrrrrrr}
\toprule
Dataset & Measure & pdQ\_dens & pdQ\_wqr & MLE & QME & MGE & Starship \\
\midrule

Flesland, Summer
& JSD
& \textbf{0.000282}
& 0.000342
& 0.000373
& 0.005071
& 0.000567
& 0.000432 \\

& Time (s)
& 0.152
& 0.045
& 2.362
& 0.254
& 3.455
& 55.037 \\

Hovden--Lundane, Winter
& JSD
& \textbf{0.002455}
& 0.002706
& 0.003549
& 0.030174
& 0.014466
& 0.003558 \\

& Time (s)
& 0.134
& 0.059
& 1.524
& 0.119
& 1.816
& 23.344 \\

Lyng{\o}r fyr, Winter
& JSD
& \textbf{0.000110}
& 0.000138
& 0.000216
& 0.000595
& 0.000129
& 0.000152 \\

& Time (s)
& 0.145
& 0.048
& 2.211
& 0.138
& 2.784
& 50.314 \\

Sande--Galleberg, Spring
& JSD
& 0.000132
& 0.000157
& \textbf{0.000126}
& 0.000207
& 0.000338
& 0.000132 \\

& Time (s)
& 0.115
& 0.055
& 1.866
& 0.150
& 2.056
& 35.587 \\

\bottomrule
\end{tabular}

\end{adjustbox}

\begin{tablenotes}
\scriptsize
\item \textit{Notes:} Computation times are mean elapsed times over five runs. Bold values indicate the smallest JSD for each dataset.
\end{tablenotes}

\end{threeparttable}
\end{table}

Table~\ref{tab:dtr_comparison} shows that pdQ\_dens gives the smallest JSD for three of the four datasets. For Sande-Galleberg, MLE performs slightly better. The pdQ\_wqr method also provides competitive density fits across the four datasets. Overall, these results are consistent with the strong density-fitting performance of pdQ\_dens observed in the simulation study.

The proposed pdQ estimators are also generally faster than the existing methods. The pdQ\_wqr method has the lowest computation time for all four datasets, while pdQ\_dens is also substantially faster than MLE, MGE, and Starship. Among the existing methods, Starship is the slowest despite giving similar JSD values in several cases.

These results show that the proposed estimators provide a useful combination of density-fitting accuracy and computational efficiency. The computational advantage becomes particularly important when repeated fitting is required. We illustrate this next through rolling-window estimation and bootstrap inference.

\subsection{Rolling distributional modelling of continuous glucose monitoring data}
\label{sec:cgm}

In this application, we consider continuous glucose monitoring (CGM) data to illustrate the use of the FPLD in a dynamic setting. CGM devices record glucose measurements at frequent intervals, and the underlying glucose distribution may change over time. Rolling-window analysis requires repeated distributional fits and therefore provides a useful application for the proposed estimators.

\subsubsection{Rolling density estimation and time in range}
\label{sec:cgm_rolling}

For this application, we use Dexcom Clarity CGM data from \cite{vigers2019cgmanalysis}, with glucose measurements recorded approximately every 5 minutes. The cleaned data are publicly available through Figshare at \url{https://figshare.com/articles/dataset/Cleaned_Files/8269736}. We consider data from a single individual and fit the distributions using a rolling window of 500 observations, moving forward by 10 observations at each step. The FPLD is fitted using the proposed pdQ\_wqr estimator and compared with the Normal, Lognormal, and Gamma distributions. We use pdQ\_wqr because of its low computational cost, which is particularly useful for repeated fitting. The Gamma distribution is included because previous research has shown its usefulness for modelling CGM data \citep{xu2024time}.

\begin{figure}[h!t]
    \centering
    \includegraphics[scale=0.2]{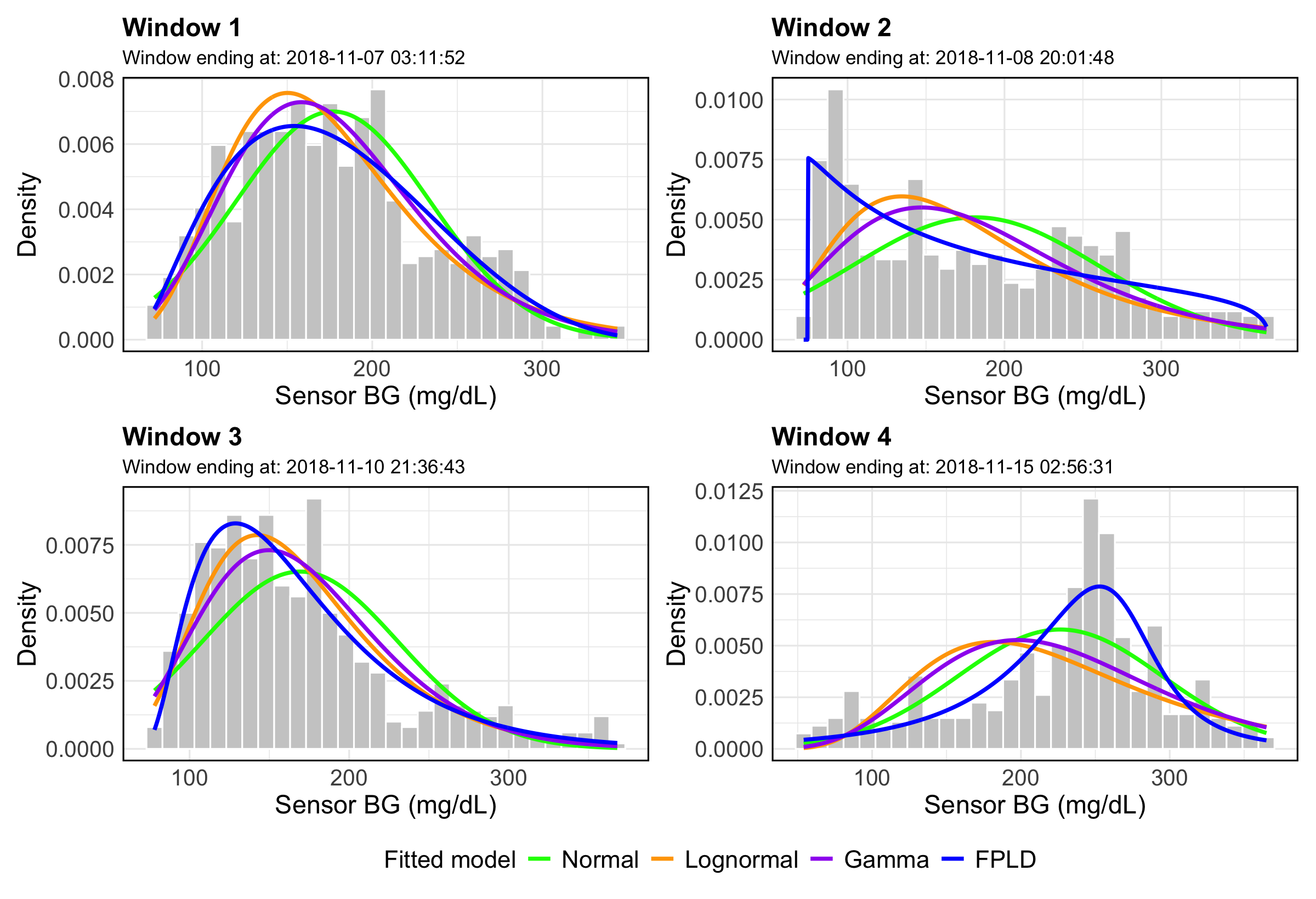}
    \caption{Rolling-window density fits for four selected CGM windows using
    the FPLD fitted by pdQ\_wqr and the Normal, Lognormal, and Gamma
    distributions. Grey histograms represent the observed glucose data and
    the curves show the fitted densities.}
    \label{fig:rolling_density_windows}
\end{figure}

Figure~\ref{fig:rolling_density_windows} shows the fitted densities for four
selected windows representing different distributional behaviours observed
throughout the CGM series. The empirical distributions vary considerably
between windows, including changes in skewness and tail behaviour. While the Normal, Lognormal, and Gamma distributions provide reasonable fits in some windows, the FPLD offers greater flexibility in capturing the different shapes observed in the data.

An advantage of modelling the complete glucose distribution is that
clinically relevant probabilities can be obtained directly from the fitted
model. One such measure is time in range (TIR), defined as the proportion of
glucose values within the recommended target interval of 70--180 mg/dL. TIR
is an important measure of short-term glycaemic control
\citep{gabbay2020time,beck2019validation}. For each rolling window, we estimate
\[
    \Pr(70 \leq X \leq 180)
\]
using the fitted FPLD, Normal, Lognormal, and Gamma distributions and compare
these estimates with the corresponding empirical proportion.

\begin{figure}[h!t]
    \centering
    \includegraphics[scale=0.2]{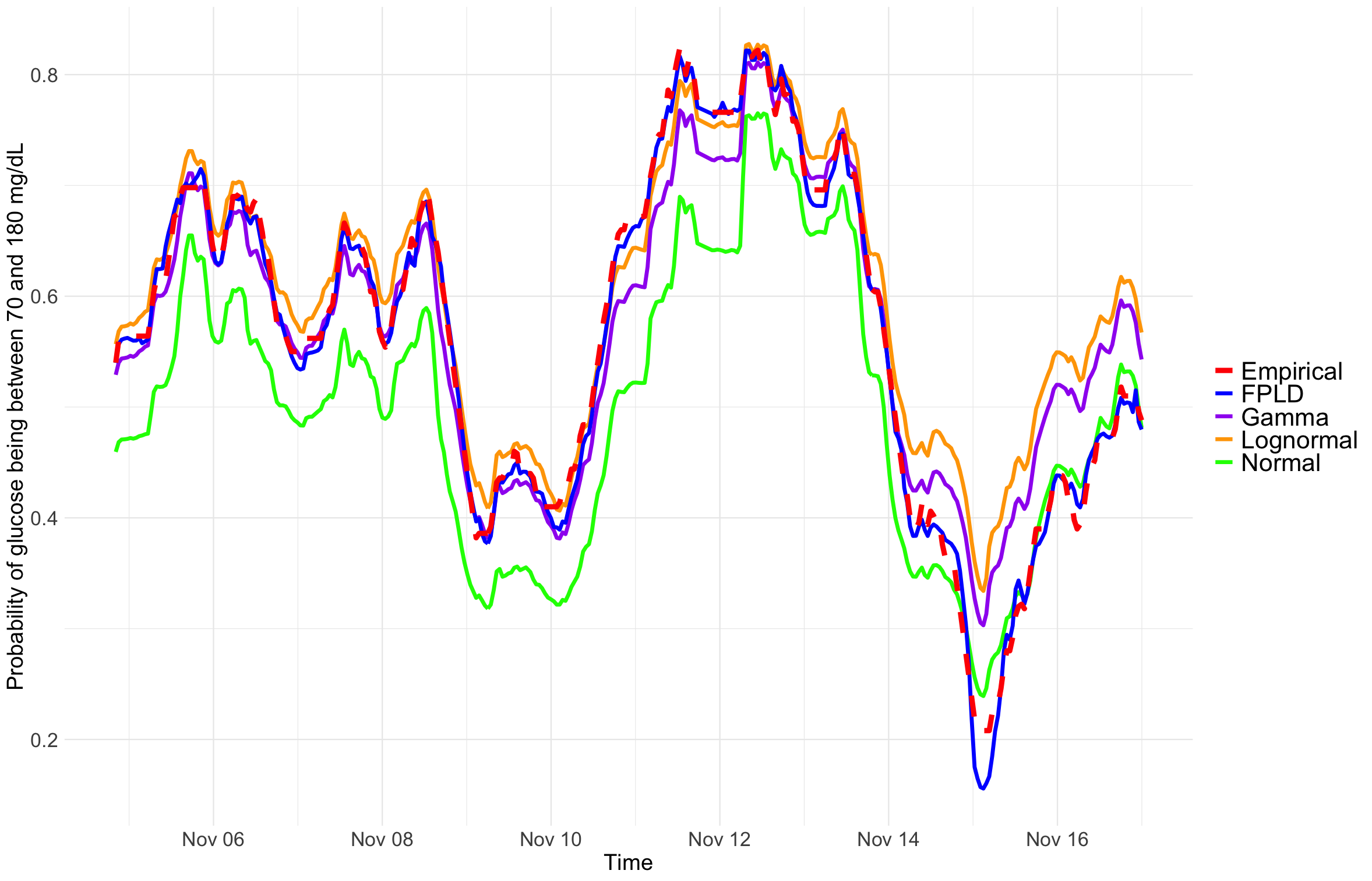}
    \caption{Rolling estimates of time in range (TIR), defined as the probability of glucose values lying between 70 and 180 mg/dL. Estimates from the FPLD (pdQ\_wqr), Normal, Lognormal, and Gamma distributions are based on rolling windows of 500 observations with a step size of 10 observations. The corresponding empirical rolling TIR is shown for comparison.}
    \label{fig:TIR}
\end{figure}

Figure~\ref{fig:TIR} shows the rolling TIR estimates obtained from the fitted distributions. The FPLD generally follows the empirical TIR pattern closely over time, while differences between the fitted distributions become more apparent as the glucose distribution changes. These results illustrate how rolling FPLD estimation can be used to monitor changes in the glucose distribution and estimate clinically relevant measures such as TIR.

The rolling analysis also highlights the importance of computational efficiency when repeated fitting is required. With a step size of 10 observations, the FPLD must be re-estimated frequently throughout the CGM series. As shown in the simulation study and DTR application, pdQ\_wqr provides competitive density fits at a low computational cost. This makes it particularly suitable for rolling-window applications requiring frequent distributional updates.

\subsubsection{Characterising changes in distributional shape}

Parametric fitting offers several advantages over non-parametric estimation methods, such as kernel density estimation. These include faster convergence \citep[e.g., p.~309,][]{wasserman}, greater stability in the presence of volatility, and improved interpretability. Here, we focus on the latter and show that the greater flexibility of the FPLD provides additional insight into this data set compared with the gamma distribution.  Specifically, we consider the estimated FPLD densities of glucose levels at 347 rolling windows of CGM data to assess changes in the shape of the glucose profile. To do so, we use the concepts of \textit{depth densities} and \textit{bagplots}, both of which are explained when the results are presented.

\begin{figure}[ht]
    \centering
    \includegraphics[width=0.85\linewidth]{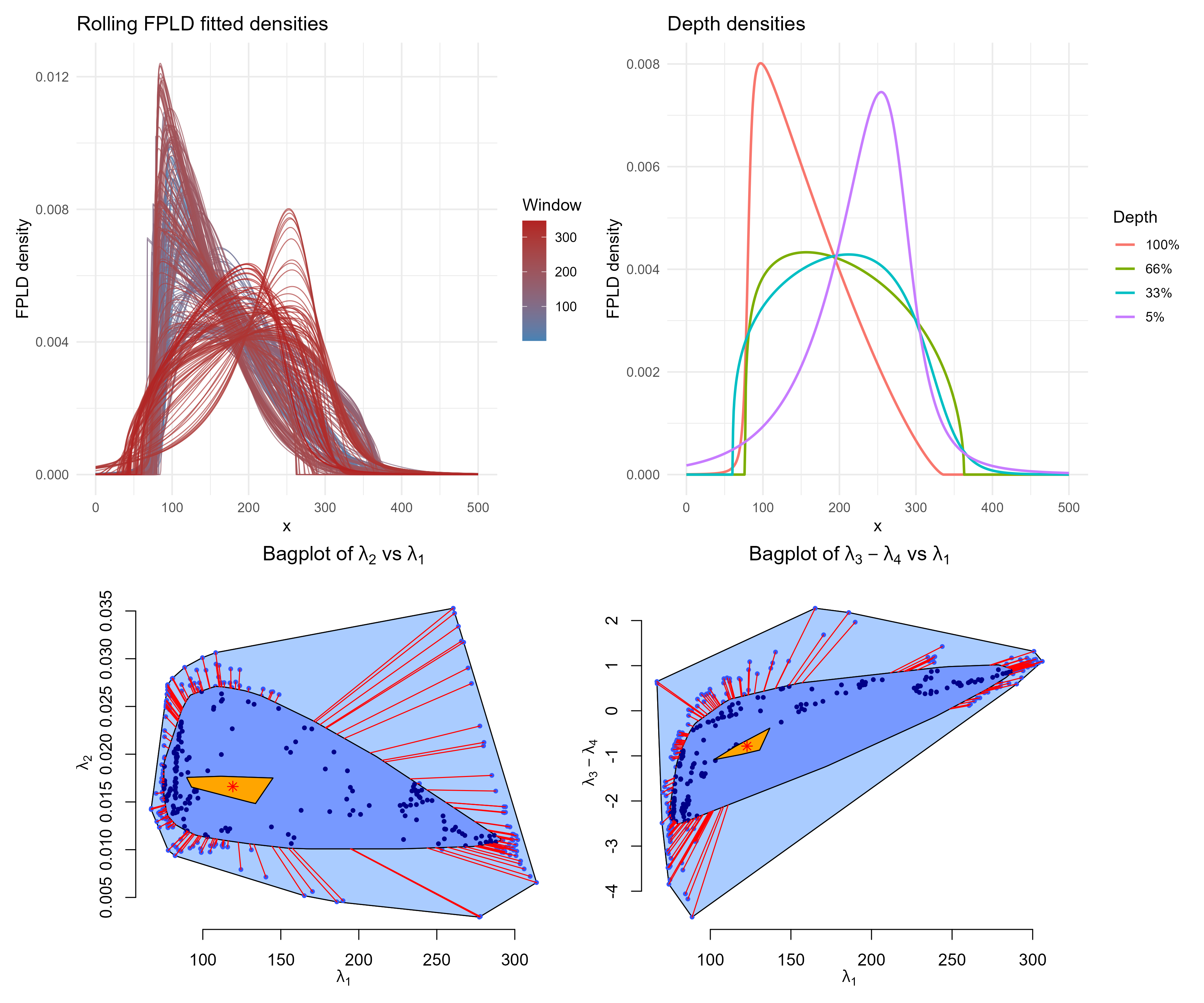}
    \caption{Estimated FPLD densities for the rolling window of fits for the glucose data (top left), the median fitted density as well as densities of different depth(top right) and bagplots of the estimated $\lambda_1$s versus the $\lambda_2$ estimates (bottom left) and estimated $\lambda_3-\lambda_4$ and estimated $\lambda_1$s (bottom right).}
    \label{fig:GLD_densities_bag_plots}
\end{figure}

Suppose that we have fitted $L$ FPLD densities across the $L$ windows and let $\widehat{\boldsymbol{\lambda}}_l=[\widehat{\lambda}_{l1},\ldots,\widehat{\lambda}_{l5}]^\top$ denote the FPLD parameter estimates for the $l$th window.  Each of these densities are depicted in Figure \ref{fig:GLD_densities_bag_plots} (top left).  We see that the estimated densities vary greatly in location, spread and shape, with three different shape profiles evident.  We first consider the geometric median (also called the spatial median) which is defined as the set of points that minimises the sum of Euclidean distances between the data rows.  In terms of our FPLD parameter estimates, the geometric median is given as 
\begin{equation}
    \widehat{\mathbf{m}}_\lambda = \underset{\mathbf{x}\in \mathbb{R}^5}{\arg\min}\sum^L_{l=1}\|\widehat{\boldsymbol{\lambda}}_l-\mathbf{x}\|
\end{equation}
and we let the median density be the FPLD density with parameter estimates equal to the values in $ \widehat{\mathbf{m}}_\lambda$.  We calculated $ \widehat{\mathbf{m}}_\lambda$ using the Weiszfeld algorithm, specifically the \texttt{Weiszfeld} function in the \texttt{Gmedian} package \citep{Gmedian}.

When we have many densities fitted and plotted, it may be difficult to determine what is a `typical' density and what is evident but not as prominent.  Therefore, we now consider the notion of data depth \citep{zuo2000general_depth} and with an emphasis of depth defined in terms of the Mahalanobis distance from the Geometric median.  Let the population Mahalanobis distance be defined as $D(\mathbf{X};\boldsymbol{\mu}, \boldsymbol{\Sigma})=\sqrt{(\mathbf{x}-\boldsymbol{\mu})^\top\boldsymbol{\Sigma}^{-1}(\mathbf{x}-\boldsymbol{\mu})}$ where $\boldsymbol{\mu}$ is a measure of central location (usually the mean) and $\boldsymbol{\Sigma}$ is a positive definite covariance matrix.  This is interpreted as a scale invariant distance of $\mathbf{x}$ from $\boldsymbol{\mu}$.  

Depth has a reciprocal relationship with distance, with depth being greater for data closer to the central point.  Define the depth of a vector $\mathbf{x}$, relative to $\boldsymbol{\mu}$ and $\boldsymbol{\Sigma}$, as
$\text{De}(\mathbf{X};\boldsymbol{\mu}, \boldsymbol{\Sigma}) = \left\{1+D^2(\mathbf{X};\boldsymbol{\mu}, \boldsymbol{\Sigma})\right\}^{-1}$.  Since we have a median set of FPLD parameters, $\widehat{\mathbf{m}}_\lambda$, we are interested in other densities at different depths relative to that median density.  Let the $l$th sample depth be
\begin{equation}
d_l = \frac{1}{1+D^2(\widehat{\boldsymbol{\lambda}}_l;\widehat{\mathbf{m}}_\lambda, \mathbf{S})}    
\end{equation}
where $\mathbf{S}$ is the sample covariance matrix of $\widehat{\boldsymbol{\lambda}}_1,\ldots,\widehat{\boldsymbol{\lambda}}_L$.  We have used the sample covariance although robust estimators of $\boldsymbol{\Sigma}$ are also often used \citep[e.g.][]{rousseeuw2011robust}.  We calculated the $d_l$s using the \texttt{depthMah} function from the \texttt{DepthProc} package \citep{DepthProc}.  We then obtained quantile estimates of these to obtain a relative measure of depth where, for example, the observed density with approximately 100\% depth is the density whose $\widehat{\boldsymbol{\lambda}}$ is closest to $\widehat{\mathbf{m}}_\lambda$, for the estimated density with 66\% depth, this is the density whose $\widehat{\boldsymbol{\lambda}}$ is closer to $\widehat{\mathbf{m}}_\lambda$ than approximately two thirds of the other densities, and so on. 
    
In the top-right plot of Figure \ref{fig:GLD_densities_bag_plots}, we provide the density with approximate 100\% depth (red line, exhibiting clear right skew), the 66\% and 33\% approximate depth densities (less skew and more centralised), and a density on the boundary of those further away from the median (5\%, blue line. clear left skew).  These depth densities help us to depict `typical' shapes in the CGM data over time with the 100\% density being the observed density closest to the median.  The variance in shapes is vast, with the 5\% depth density indicating that left skew is possible, although less likely than the other shapes.

To examine the relationship between parameters with reference to the median, we used a bagplot \citep{bagplot} which is a bivariate extension of the boxplot.  Bagplots use a different definition of depth, namely halfspace depth, but notionally depth is similar.  In the bottom-left plot of Figure \ref{fig:GLD_densities_bag_plots} we provide a bagplot of the estimated $\lambda_1$s and estimated $\lambda_2$s.  The orange `asterix' is the halfspace median, and the orange hull around the median is the area of greatest depth.  The dark blue `bag' contains 50\% of the data, and the lighter blue is the inflated bag that can be used to detect outliers.  We can see within the bag containing the combination of estimates which have greater depth, that when the $\lambda_1$ estimate (location) is larger, there is less variation in the inverse scale parameter estimate. However, the densities in the outer bag can have a lot of variation in scale for larger location values.  This is evident in the density plots, where lower density peaks are evident to the right (larger location) values, with more mass in the tails. Although there is also a relationship with the $\lambda_5$ parameters, the direction of skew is largely determined by the difference $\lambda_3-\lambda_4$.  In the bottom-right plot we can see a clear relationship between location, $\lambda_1$, and $\lambda_3-\lambda_4$ with the skewness changing as, including in direction, as the estimated $\lambda_1$s increase.  This highlights the differences in shape across densities, even for those with greater depth relative to the median.      

\begin{figure}[ht]
    \centering
    \includegraphics[width=0.85\linewidth]{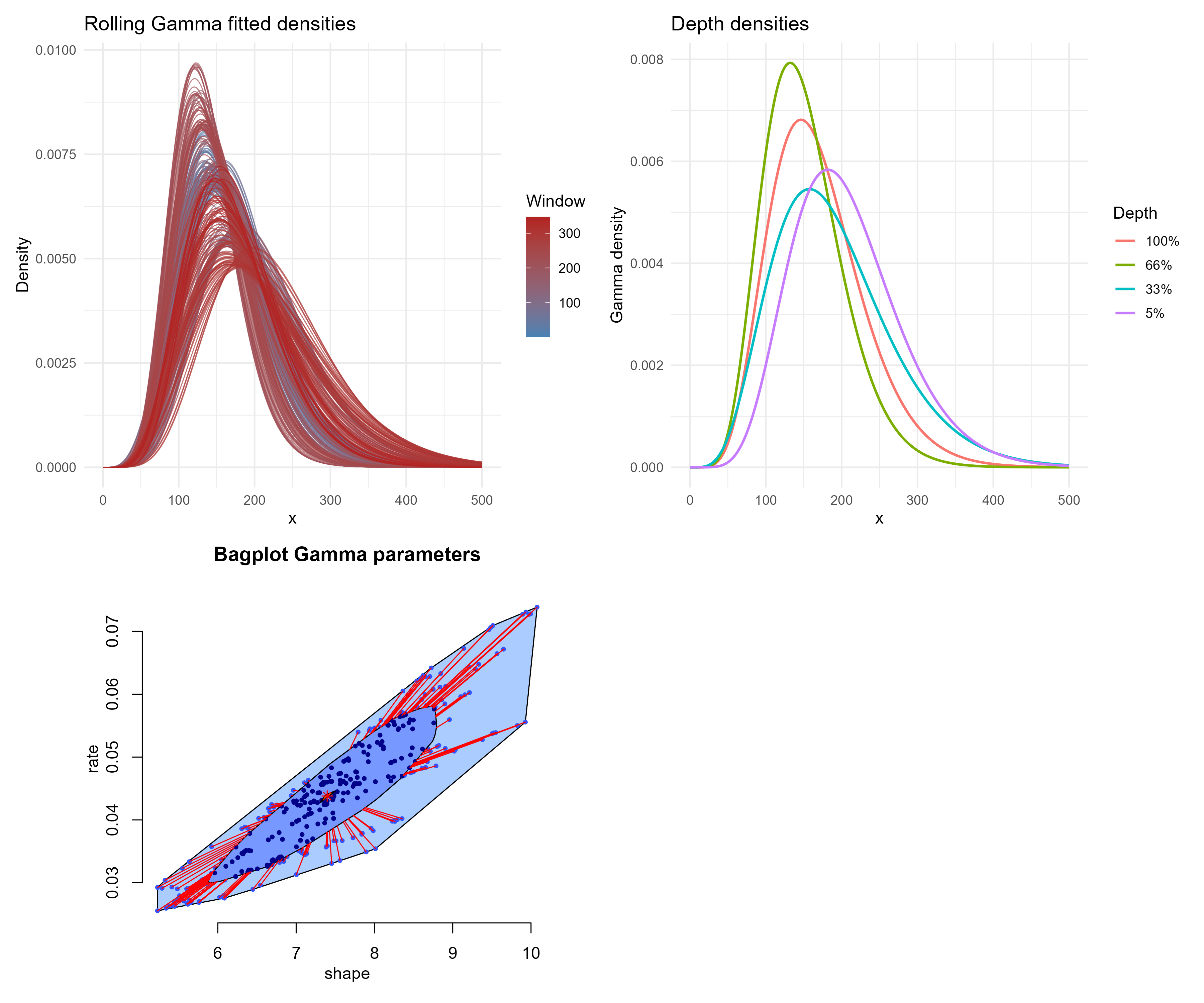}
    \caption{Estimated gamma densities for the rolling window of fits for the glucose data (top left), the median fitted density as well as densities of different depth (top right) and bagplots of the estimated shape and rate estimates (bottom left).}
    \label{fig:gamma_densities_bag_plots}
\end{figure}

In contrast, the fitted gamma densities are not as informative as those achieved using the FPLD.  The gamma distribution is limited in shape and, while differences are observed across the rolling windows (see the top-left plot of Figure \ref{fig:gamma_densities_bag_plots}), the gamma distribution cannot model left skew and can be poor at approximating symmetry.  The shape similarities in the median and depth densities is also evident (top-right plot), and even the 5\% depth density (relatively far from the median), is very similar to the 33\% depth density.  The bagplot of the shape and rate estimates (bottom-left plot) is more condensed within the inner 50\% bag.  These gamma plots show the stability of the data patterns in rolling windows, which is not the case as evidenced by the histogram snapshots in Figure \ref{fig:rolling_density_windows}, and the FPLD findings in Figure \ref{fig:GLD_densities_bag_plots}.

\subsection{Bootstrap confidence bands for estimated densities}
\label{sec:bootstrap}

The applications considered so far have focused primarily on estimation of
the FPLD. In many applications, however, it is also important to quantify
uncertainty associated with an estimated density. Confidence bands provide
a useful way to assess this uncertainty over the domain of the distribution,
either pointwise at individual values or simultaneously over a specified
range.

For the FPLD, analytical uncertainty quantification can be challenging
because the fitted density is a nonlinear function of five estimated
parameters. An asymptotic approach based on the covariance matrix of the
parameter estimates would require this uncertainty to be propagated through
the fitted density, for example using the delta method. Such approximations
may be difficult to implement, particularly when uncertainty is required
over the entire fitted density rather than at a single point.

Bootstrap methods provide a convenient alternative for quantifying
uncertainty in the fitted density without requiring an explicit analytical
variance expression \citep{efron1987better,mcdonald2021review}. Both
pointwise confidence intervals and simultaneous confidence bands can be
constructed from the resulting bootstrap distributions. Bootstrap inference,
however, requires the FPLD to be re-estimated for a large number of
resampled data sets. Consequently, computationally efficient estimation is
particularly important in this setting, making the proposed pdQ\_wqr estimator
well suited to repeated bootstrap fitting. Details of the bootstrap
procedures used to construct the pointwise and simultaneous confidence bands
are provided in Appendix~\ref{app:bootstrap}.

\begin{figure}[H]
    \centering
    \includegraphics[scale=0.3]{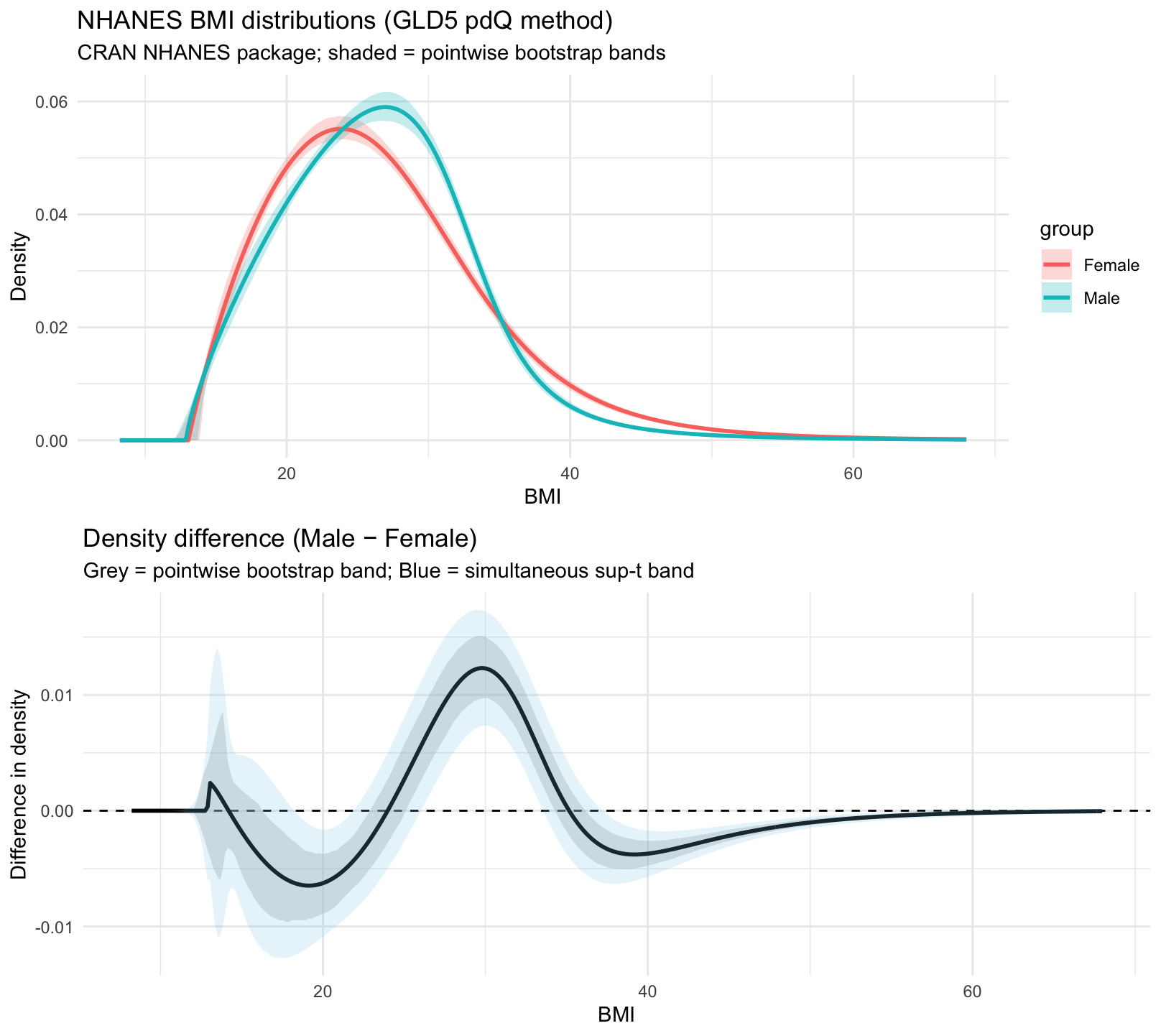}
    \caption{Bootstrap inference for FPLD densities fitted to NHANES BMI
    data using pdQ-wqr. The upper panel shows the estimated female and male
    BMI densities with pointwise bootstrap confidence bands. The lower panel
    shows the estimated difference between the male and female densities,
    together with pointwise and simultaneous bootstrap confidence bands.
    The overall computation time was 116.84 seconds.}
    \label{fig:bmi_bootstrap}
\end{figure}

To illustrate the approach, we use body mass index (BMI) measurements
from female and male participants in the 2009--2012 cycles of the US
National Health and Nutrition Examination Survey (NHANES), as provided
in the NHANES R package \citep{NHANES}. The FPLD is fitted
separately to the two groups using pdQ\_wqr, and bootstrap resampling is used
to quantify uncertainty in the fitted densities.
Figure~\ref{fig:bmi_bootstrap} shows the estimated female and male BMI
densities together with pointwise bootstrap confidence bands. The lower
panel shows the difference between the male and female fitted densities,
together with pointwise and simultaneous bootstrap confidence bands.

The fitted densities indicate differences in the distribution of BMI between
the two groups, while the bootstrap bands quantify the uncertainty associated
with the estimated densities and their difference across the range of BMI
values. More importantly in the context of the present study, this example
demonstrates that bootstrap inference for FPLD densities can be implemented
using repeated pdQ\_wqr estimation at relatively low computational cost. The
complete analysis, including the repeated fitting required to construct the
confidence bands, was completed in 116.84 seconds.

\section{Software Availability}
\label{sec:software}

To facilitate practical use of the proposed estimation approach, an R package, FPLDpdQ, has been developed for fitting the FPLD to observed data using the proposed two-stage pdQ method. The package implements pdQ-based shape parameter estimation followed by several alternative methods for estimating the location and scale parameters, including quartile matching, density matching, and quantile regression. The package is publicly available on GitHub at \url{https://github.com/shenal-dkumara/FPLDpdQ} and can be installed directly from the repository.

In addition, an interactive Shiny application was developed to approximate selected probability distributions using the FPLD. The application allows users to specify a target distribution and its parameters and obtain the corresponding FPLD parameter estimates using the pdQ approach. Location and scale parameters can subsequently be estimated using either distributional least squares or quartile matching. The resulting approximation can be examined graphically through density and pdQ plots, with Hellinger distance and Jensen–Shannon divergence also reported as measures of approximation accuracy.  The application is publicly available at \url{https://shenal.shinyapps.io/FPLD-approximation/} and the authors can be contacted if there are any issues.

\section{Discussion and conclusions}
\label{sec:discussion}

In this paper, we present new computationally efficient estimators for the FPLD that compare favourably with existing estimators at a fraction of the computational cost. The three shape parameters are isolated using the location- and scale-invariant pdQ \citep{staudte2017, dedduwakumara2021efficient} and then estimated by minimising the Hellinger distance between the empirical and FPLD pdQs. Several estimators of the location and scale parameters, conditional on the shape estimates, were considered. We recommend the density minimisation method (pdQ\_dens) or the weighted quantile regression (pdQ\_wqr) for estimating location and scale. Our simulation studies showed that pdQ\_dens often produced slightly better density estimates, whereas pdQ\_wqr was computationally faster. Both methods performed well compared with existing estimators, such as Starship and MLE, while requiring substantially less computation time.

We also presented several applications highlighting the usefulness of the FPLD as a highly flexible parametric density estimator compared with commonly used distributions that may miss important features, particularly with respect to skewness and tail behaviour. We further highlighted the importance of fast computation when repeated estimation is required, such as in rolling-window estimation and bootstrapping. An R package implementing the proposed FPLD estimators is publicly available on GitHub, and a web application is available for selecting FPLD parameters to approximate several well-known distributions (see Section \ref{sec:software} for details).

\appendix

\section{Additional results for the pdQ estimators}
\label{app:simulation_results}

For completeness, Table~\ref{tab:JSD_pdQ_additional} reports the simulation results for the additional pdQ estimators not presented in the main comparison. These include the Unweighted quantile regression (pdQ\_qr with $\tau=0.5$), quartile-matching (pdQ\_qm) and method-of-moments (pdQ\_mom) estimators. Results for pdQ\_dens and pdQ\_wqr($\tau=0.5$), together with the existing estimation methods, are reported in Table~\ref{tab:JSD} of the main text.  

\begin{table}[H]
\centering
\caption{Average Jensen--Shannon divergence (JSD) between the empirical
density and the fitted FPLD density over 500 simulation replications for
the additional pdQ estimators.}
\label{tab:JSD_pdQ_additional}

\begin{threeparttable}

\begin{adjustbox}{max width=\textwidth}
\footnotesize

\begin{tabular}{ccccc}
\toprule
Distribution & $n$ & pdQ\_qr & pdQ\_qm & pdQ\_mom \\
\midrule

Normal(0,1)
& 100  & 0.0010 & 0.0018 & 0.0343 \\
& 250  & 0.0004 & 0.0006 & 0.0012 \\
& 500  & 0.0003 & 0.0004 & 0.0004 \\
& 1000 & 0.0002 & 0.0003 & 0.0002 \\

\midrule

Lognormal(0,1)
& 100  & 0.0340 & 0.0198 & 0.3398  \\
& 250  & 0.0130 & 0.0093 & 0.2305 \\
& 500  & 0.0073 & 0.0063 & 0.1831  \\
& 1000 & 0.0044 & 0.0041 & 0.1396 \\

\midrule

Exponential(1)
& 100  & 0.0275 & 0.0337 & 0.1921  \\
& 250  & 0.0218 & 0.0224 & 0.1556 \\
& 500  & 0.0145 & 0.0160 & 0.1199 \\
& 1000 & 0.0109 & 0.0116 & 0.0765 \\

\midrule

Dagum
& 100  & 0.0079 & 0.0061 & 0.1266 \\
& 250  & 0.0013 & 0.0018  & 0.0611 \\
& 500  & 0.0008 & 0.0012 & 0.0345 \\
& 1000 & 0.0004 & 0.0005 & 0.0149 \\

\midrule

Singh--Maddala
& 100  & 0.0041 & 0.0077 & 0.0752 \\
& 250  & 0.0016 & 0.0023 & 0.0127 \\
& 500  & 0.0010 & 0.0015 & 0.0044 \\
& 1000 & 0.0006 & 0.0007 & 0.0014 \\

\midrule

$\chi^2_{5}$
& 100  & 0.0032 & 0.0055 & 0.0733 \\
& 250  & 0.0014 & 0.0022 & 0.0079 \\
& 500  & 0.0009 & 0.0012 & 0.0021 \\
& 1000 & 0.0005 & 0.0006 & 0.0010 \\

\midrule

Beta(2,3)
& 100  & 0.0020 & 0.0041 & 0.0381 \\
& 250  & 0.0008 & 0.0015 & 0.0012 \\
& 500  & 0.0005 & 0.0008 & 0.0005 \\
& 1000 & 0.0003 & 0.0005 & 0.0003 \\

\bottomrule
\end{tabular}

\end{adjustbox}
\end{threeparttable}
\end{table}

\section{Bootstrap algorithms}
\label{app:bootstrap}

The bootstrap procedures used to construct the pointwise and simultaneous
confidence bands presented in Section~\ref{sec:bootstrap} are described below. Algorithm~1
provides pointwise confidence bands for a single fitted FPLD density, while
Algorithm~2 extends this procedure to simultaneous confidence bands.
Algorithm~3 describes the corresponding procedure for the difference between
two independently fitted FPLD densities.

Throughout, let $\mathcal{X}_n=\{x_1,\ldots,x_{n}\}$ denote a sample of $n$ observations.  In Algorithm \ref{alg:boot-density} we provide details of the point-wise $(1-\alpha)100$\% bootstrapping process for the FPLD density.   We repeatedly sample $n$ observations, with replacement, from $\mathcal{X}_n$ and estimated the FPLD density for each sample. In total, there are $B$ estimated FPLD densities and where $B$ is typically chosen to be large (e.g. 500 or 1000 are common choices).  Given the large number of estimates required, the good estimator performance, and small computation times of the pdQ estimators, make the pdQ estimator a viable choice for this process.

\begin{algorithm}[ht]
\caption{Bootstrap pointwise $1-\alpha$ confidence bands for one FPLD density}
\label{alg:boot-density}
\begin{algorithmic}[1]

\State Fit the FPLD to $\mathbf{x}$ and obtain parameter estimate $\hat{\boldsymbol{\lambda}}$
\State For grid $\mathcal{G}=\{g_1,\ldots,g_r\}$, compute fitted density
\[
\hat f(g_j) = f(g_j;\hat{\boldsymbol{\lambda}}),
\qquad j=1,\ldots,r.
\]

\For{$b=1,\ldots,B$}
    \State Draw a bootstrap sample $\mathbf{x}_{(b)}$ by sampling $n$ observations from $\mathbf{x}$ with replacement
    \State Fit FPLD to $\mathbf{x}_{(b)}$ and obtain $\hat{\boldsymbol{\lambda}}_{(b)}$
    \State Compute the bootstrap density
    \[
    \hat f_{(b)}(g_j)
    =
    f(g_j;\hat{\boldsymbol{\lambda}}_{(b)}),
    \qquad j=1,\ldots,m.
    \]
\EndFor

\For{$j=1,\ldots,r$}
    \State Compute the pointwise confidence band
    \[
    \ell(g_j)
    =
    q
    \left(
    \hat f_{(1)}(g_j),\ldots,\hat f_{(B)}(g_j); \alpha/2
    \right), \quad     u(g_j)
    =
    q
    \left(
    \hat f_{(1)}(g_j),\ldots,\hat f_{(B)}(g_j); 1-\alpha/2
    \right)
    \]

\EndFor

\State \Return $\{\hat f(g_j),\ell(g_j),u(g_j)\}_{j=1}^m$
\end{algorithmic}
\end{algorithm}

\begin{algorithm}[ht]
\caption{Bootstrap simultaneous $1-\alpha$ confidence bands for one FPLD density}
\label{alg:sim_boot-density-diff}
\begin{algorithmic}[1]

\State \text{Assume:} Following Algorithm \ref{alg:boot-density}

    \For{$j=1,\ldots,r$}
        \State Estimate the bootstrap SE
        \[
        \operatorname{SE}(g_j)
        =
        \operatorname{sd}
        \left(
        \hat{f}_{(1)}(g_j),\ldots,\hat{f}_{(B)}(g_j)
        \right).
        \]
    \EndFor

    \For{$b=1,\ldots,B$}
        \State Compute the standardized bootstrap $z$-scores
        \[
        Z_{(b)}(g_j)
        =
        \frac{
        \hat f_{(b)}(g_j)-\hat f(g_j)
        }{
        \operatorname{SE}(g_j)
        },
        \qquad j=1,\ldots,m.
        \]
        \State Compute
        \[
        T_{(b)}=\max\left\{
        \left|Z^{*(b)}(g_1)\right|,\ldots,\left|Z^{*(b)}(g_r)\right|\right\}.
        \]
    \EndFor

    \State Let
    \[
    q_{1-\alpha}
    =
    q
    \left(
    T_{(1)},\ldots,T_{(B)};1-\alpha/2
    \right).
    \]

    \For{$j=1,\ldots,m$}
        \State Compute the simultaneous confidence band
        \[
        \ell_s(g_j)
        =
        \hat f(g_j)
        -
        q_{1-\alpha}\operatorname{SE}(g_j),\quad         u_s(g_j)
        =
        \hat f(g_j)
        +
        q_{1-\alpha}\operatorname{SE}(g_j).
        \]
    \EndFor

\State \Return $\{\hat{f}(g_j),\ell_s(g_j),u_s(g_j)\}_{j=1}^m$
\end{algorithmic}
\end{algorithm}

\begin{algorithm}[ht]
\caption{Bootstrap $1-\alpha$ confidence bands for the difference of two densities}
\label{alg:boot-density-diff}
\begin{algorithmic}[1]

\State Fit FPLDs to $\mathbf{x}$ and $\mathbf{y}$ to get $\hat{\boldsymbol{\lambda}}_x$ and $\hat{\boldsymbol{\lambda}}_y$
\State Compute
\[
\hat f_x(g_j)=f(g_j;\hat{\boldsymbol{\lambda}}_x),
\;
\hat f_y(g_j)=f(g_j;\hat{\boldsymbol{\lambda}}_y),\; \widehat{\Delta}(g_j)=\hat f_1(g_j)-\hat f_2(g_j),
\; j=1,\ldots,m.
\]

\For{$b=1,\ldots,B$}
    \State Draw bootstrap samples $\mathbf{x}_{(b)}$and $\mathbf{y}_{(b)}$ with replacement from $\mathbf{x}$ and $\mathbf{y}$
    \State Fit FPLDs to $\mathbf{x}_{(b)}$ and $\mathbf{y}_{(b)}$ to get $\hat{\boldsymbol{\lambda}}_{x,(b)}$ and $\hat{\boldsymbol{\lambda}}_{y,(b)}$
    \State For $j=1,\ldots,m$ compute bootstrap densities and difference of densities
    \[
    \hat f_{x,(b)}(g_j)
    =
    f(g_j;\hat{\boldsymbol{\lambda}}_{x,(b)}),
    \;
    \hat f_{y,(b)}(g_j)
    =
    f(g_j;\hat{\boldsymbol{\lambda}}_{y,(b)}), \;  \widehat\Delta_{(b)}(g_j)
    =
    \hat f_{x,(b)}(g_j)
    -
    \hat f_{y,(b)}(g_j)
    \]
   
\EndFor

\For{$j=1,\ldots,m$}
    \State Let
    \[
    \ell(g_j)
    =
    q
    \left(
    \widehat{\Delta}_{(1)}(g_j),\ldots,\widehat{\Delta}_{(B)}(g_j); \alpha/2
    \right),
    \]
    \[
    u(g_j)
    =
        q
    \left(
    \widehat{\Delta}_{(1)}(g_j),\ldots,\widehat{\Delta}_{(B)}(g_j); 1-\alpha/2
    \right).
    \]
\EndFor

\State \Return $\{\widehat{\Delta}(g_j),\ell(g_j),u(g_j)\}_{j=1}^m$
\end{algorithmic}
\end{algorithm}

\clearpage
\bibliographystyle{authordate4}
\bibliography{ref}

\end{document}